\documentclass[superscriptaddress,aps,letterpaper,10pt,twocolumn,floats,showpacs,amsmath,amsfonts,amssymb,pre,nofootinbib]{revtex4-2}
\usepackage{graphicx}% Include figure files
\usepackage[colorlinks=true,citecolor=blue]{hyperref}
\usepackage[caption=false]{subfig}
\usepackage{pstricks}
\usepackage{pst-node}
\usepackage{amsmath}
\usepackage{amsbsy}
\usepackage{verbatim}
\usepackage{dsfont}
\usepackage{MnSymbol}

\newcommand{\Var}{\text{Var}}
\DeclareMathOperator*{\argmin}{arg\,min}
\newcommand{\re}{\text{Re}}

\begin{document}
	
	\title{Critical heat current fluctuations in Curie-Weiss model in and out of equilibrium}% Force line breaks with \\
	
	\author{Krzysztof Ptaszy\'{n}ski}
\email{krzysztof.ptaszynski@ifmpan.poznan.pl}

\affiliation{Complex Systems and Statistical Mechanics, Department of Physics and Materials Science, University of Luxembourg, 30 Avenue des Hauts-Fourneaux, L-4362 Esch-sur-Alzette, Luxembourg}
	\affiliation{Institute of Molecular Physics, Polish Academy of Sciences, Mariana Smoluchowskiego 17, 60-179 Pozna\'{n}, Poland}
	
	\author{Massimiliano Esposito}
\affiliation{Complex Systems and Statistical Mechanics, Department of Physics and Materials Science, University of Luxembourg, 30 Avenue des Hauts-Fourneaux, L-4362 Esch-sur-Alzette, Luxembourg}
	
	\date{\today}
	
	\begin{abstract}
In some models of nonequilibrium phase transitions, fluctuations of the analyzed currents have been observed to diverge with system size. To assess whether this behavior is universal across phase transitions, we examined heat current fluctuations in the Curie-Weiss model, a paradigmatic model of the paramagnetic-ferromagnetic phase transition, coupled to two thermal baths. This model exhibits phase transitions driven by both the temperature and the magnetic field. We find that at the temperature-driven phase transition, the heat current noise consists of two contributions: the equilibrium part, which vanishes with system size, and the nonequilibrium part, which diverges with system size. For small temperature differences, this leads to nonmonotonic scaling of fluctuations with system size. In contrast, at the magnetic-field-driven phase transition, heat current fluctuations do not diverge when observed precisely at the phase transition point. Instead, out of equilibrium, the noise is enhanced at the magnetic field values away but close to the phase transition point, due to stochastic switching between two current values. The maximum value of noise increases exponentially with system size, while the position of this maximum shifts towards the phase transition point. Finally, on the methodological side, the paper demonstrates that current fluctuations in large systems can be effectively characterized by combining a path integral approach for macroscopic fluctuations together with an effective two-state model describing subextensive transitions between the two macroscopic states involved in the phase transition.
	\end{abstract}
	
	\maketitle

\section{Introduction}
Phase transitions, that is, abrupt changes in the system properties with a small change of some control parameter, are among the most notable manifestations of collective behavior found in nature. In recent decades, significant interest has been focused on phase transitions that occur out of equilibrium~\cite{schlogl1972chemical}. Such transitions are associated with even more complex physics than their equilibrium counterparts, as they are not governed solely by the system thermodynamics but also by its kinetics. As a result, they can lead to novel phenomena absent at equilibrium, such as phase separation with purely repulsive interactions~\cite{cates2015motility}, the emergence of limit cycles~\cite{lefever1971chemical, herpich2018collective, herpich2019universality}, or the formation of Turing patterns~\cite{prigogine1968symmetry, falasco2018information}. Considerable attention has also been paid to nonequilibrium phase transitions in open quantum systems~\cite{drummond1980quantum, carr2013nonequilibrium, fitzpatrick2017observation, rodriguez2017probing,fink2018signatures}, among others, because of their potential metrological applications~\cite{fernandez2017quantum, heugel2019quantum, ilias2022criticality}.

Most studies of equilibrium and nonequilibrium phase transitions focused on the characterization of average quantities, such as magnetization, entropy production, etc. However, a valuable insight into the dynamics and thermodynamics of physical systems can be obtained by analyzing current fluctuations. For example, they can provide information on energy dissipation (via the so-called thermodynamic uncertainty relations)~\cite{barato2015thermodynamic, pietzonka2016universal, gingrich2016dissipation}, the structure of the Markovian network in classical stochastic systems~\cite{shaevitz2005statistical, moffitt2010methods, bruderer2014inverse, barato2015skewness, ptaszynski2022bounds, gerry2023random}, or quantum coherent effects~\cite{urban2008coulomb, ho2019counting, ptaszynski2018coherence, ptaszynski2022bounds, saryal2019thermodynamic}. Recently, several studies have been concerned with the finite-size scaling behavior of current fluctuations at nonequilibrium phase transitions. They demonstrated divergent scaling of fluctuations with system size (or another parameter that plays this role). Specifically, a power-law scaling of fluctuations at continuous phase transitions has been reported for certain chemical reaction models~\cite{nguyen2018phase,remlein2024nonequilibrium}, stochastic time crystals~\cite{oberreiter2021stochastic}, and open quantum systems~\cite{kewming2022diverging} (although in the latter case an exponential scaling of fluctuations of certain currents due to quantum measurement effects was also shown). At discontinuous phase transitions, instead, one usually observes an exponential scaling of fluctuations~\cite{nguyen2020exponential,fiore2021current,kewming2022diverging,remlein2024nonequilibrium}. This behavior is related to stochastic switching between the system attractors, whose timescale increases exponentially with system size~\cite{hanggi1984bistable, hinch2005exponentially,kurchan2009equilibrium}. Such a switching has been observed in real time in open quantum systems~\cite{rodriguez2017probing, fink2018signatures}. However, we have recently found that when the discontinuous phase transition is not associated with phase coexistence, one observes rather a power-law scaling of fluctuations~\cite{ptaszynski2024finite}.

In this paper, we aim to explore whether phase transitions are always associated with divergent behavior of current fluctuations at the phase transition point. To this end, we analyze the heat current fluctuations in the Curie--Weiss model, a paradigmatic model of the paramagnetic--to--ferromagnetic phase transition~\cite{kochmanski2013curie}. The nonequilibrium driving is applied to the system by connecting it to two baths with different temperatures. We show that nonequilibrium heat current fluctuations exhibit a power-law divergence at temperature-driven continuous phase transitions, even though the equilibrium fluctuations vanish at this point. Instead, at the magnetic-field-driven transition, the fluctuations calculated exactly at the phase transition point do not diverge. However, a strong noise enhancement can be observed in close proximity to the phase transition.

The paper is organized as follows. In Sec.~\ref{sec:modmeth} we introduce the analyzed model and discuss its critical behavior. In Sec.~\ref{sec:fluctmeth} we present the methods used to calculate the current fluctuations and responses. Sec.~\ref{sec:fluctscal} contains the main results of the paper on finite-size scaling of heat current fluctuations. Finally, Sec.~\ref{sec:concl} presents conclusions that follow from our results.
 
	\section{Nonequilibrium Curie--Weiss model} \label{sec:modmeth}
 \subsection{Description of the model} \label{subsec:mod}
Let us first present the considered nonequilibrium Curie--Weiss model. It consists of $N$ spins coupled through a homogeneous all-to-all Ising interaction. The energy of a particular spin configuration reads
\begin{align}
E=-\frac{J}{2N} \sum_{i,j=1}^N  \sigma_i \sigma_j-h \sum_{i=1}^N \sigma_i \,,
\end{align}
where $J \geq 0$ is the strength of the ferromagnetic Ising interaction, and $h$ is the magnetic field. Spins $\sigma_i$ are here classical random variables taking values $\pm 1$. We further define the total magnetization of the system as $M=\sum_i \sigma_i \in \{-N,-N+2,\ldots,N\}$. Due to the all-to-all nature of the coupling, the energy of the system can be rewritten as a function of the total magnetization:
\begin{align} \label{eq:modelenergy}
	E_M=-\frac{J}{2N} M^2-hM \,.
\end{align}
Each value of the total magnetization corresponds to $\Omega_M$ possible microscopic spin configurations, where
\begin{align} \label{eq:omegam}
\Omega_M=\frac{N!}{[(N+M)/2]! [(N-M)/2]!} \,.
\end{align}

To drive the system out of equilibrium, we connect it to $d$ ideal thermal baths $\alpha \in \{1,\ldots,d \}$ with temperatures $T_\alpha$ [inverse temperatures $\beta_\alpha \equiv 1/(k_B T_\alpha)$]. Similar nonequilibrium models have been previously studied in Refs.~\cite{garrido1987stationary,blote1990critical,tamayo1994twotemperature,garrido1996complexity,vroylandt2018collective,vroylandt2020efficiency,aron2020landau, herpich2020njp,beyen2024phase}\footnote{We note that nonequilibrium driving can be also induced by an alternating magnetic field, see Refs.~\cite{tome1990dynamic,sides1998kinetic,acharyya1999nonequilibrium,baek2014nonequilibrium,fiori2024specific}.}. The dynamics of the model corresponds to a Markovian flipping of individual spins due to interaction with the thermal baths. Each flipping of spin from $1$ to $-1$ ($-1$ to $1$) changes the total magnetization as $M \rightarrow M-2$ ($M \rightarrow M+2$). To simplify the problem, we also assume that, for each microscopic configuration with total magnetization $M$, the rate of flipping of each spin with a given orientation is the same. Dynamics can then be described using the mesoscopic master equation for probabilities $p_M$ of coarse-grained mesostates with definite magnetization $M$~\cite{aron2020landau, herpich2020njp, meibohm2022finite, meibohm2023landau},
\begin{align} \label{eq:masteq}
\dot{p}_M=\sum_{\pm} \left( W_{M,M\pm 2} p_{M \pm 2} - W_{M \pm 2,M} p_M \right) \,,
\end{align}
where $W_{M,M \pm 2}$ is the transition rate from the state with magnetization $M \pm 2$ to the state with magnetization $M$. For a formal derivation of this form of the master equation, see Ref.~\cite{herpich2020njp}. Each rate can be further decomposed into components associated with individual baths $\alpha$,
\begin{align}
W_{M,M \pm 2} = \sum_{\alpha=1}^d W^\alpha_{M,M\pm 2} \,.
\end{align}
To provide consistency with thermodynamics, such components have to fulfill the local detailed balance condition~\cite{herpich2020njp, FalascoReview}
\begin{align} \label{eq:locdetbal}
	\ln \frac{W^\alpha_{M,M \pm 2}}{W^\alpha_{M \pm 2,M}}=\beta_\alpha (F^\alpha_{M \pm 2}-F^\alpha_M) \,,
\end{align}
where
\begin{align}
F_M^\alpha=E_M -\beta_\alpha^{-1} \ln \Omega_M
\end{align}
is the free energy potential with respect to the bath $\alpha$. By construction, this condition provides consistency with the laws of thermodynamics, such as the first and the second law of thermodynamics, or fluctuation theorems~\cite{herpich2020njp,FalascoReview}. Physically, it means that each bath tends to relax the system to the equilibrium state with respect to this bath. In fact, for equal bath temperatures $\beta_\alpha=\beta$, the steady state of the system is the Gibbs state $p_M \propto \exp(-\beta F_M)$.

Following Refs.~\cite{herpich2020njp,meibohm2022finite,meibohm2023landau}, we further focus on a particular model of transition rates fulfilling the condition~\eqref{eq:locdetbal}, namely, the Arrhenius rates
\begin{align}
	W_{M \pm 2,M}^\alpha=\frac{\Gamma_\alpha (N \mp M)}{2} e^{-\beta_\alpha (E_{M \pm 2}-E_M)/2} \,,
\end{align}
where $\Gamma_\alpha$ are kinetic rates describing the strength of coupling to different reservoirs. We notice that, out of equilibrium, the nonequilibrium steady state depends not only on the temperatures of the baths $T_\alpha$, but also on the details of coupling to the reservoirs, parameterized by the kinetic rates $\Gamma_\alpha$.

 \subsection{Mean field and large deviation approaches} \label{subsec:mfldev}
Our aim is now to describe the nonequilibrium phase transitions in the considered model, that occur in the thermodynamic limit $N \rightarrow \infty$. To this end, one requires methods suitable for characterizing the macroscopic behavior of the system. One possible approach is the mean field theory~\cite{garrido1987stationary,garrido1996complexity,herpich2020njp,vroylandt2018collective,vroylandt2020efficiency,aron2020landau}. Within this framework, from the mesoscopic master equation~\eqref{eq:masteq} one derives an effective deterministic equation of motion for the normalized magnetization $m \equiv M/N$. For the model considered, it reads
 \begin{align} \label{eq:meanfield}
 \dot{m}=2 \left[w_+(m)-w_-(m) \right] \,,
 \end{align}
where $w_\pm(m) =\sum_{\alpha=1}^d w_\pm^\alpha(m)$ and \begin{align}
w_\pm^\alpha (m) =\lim_{N \rightarrow \infty} \frac{W_{M \pm 2,M}^\alpha}{N}= \frac{\Gamma_\alpha \left(1 \mp m \right)}{2} e^{\pm \beta_\alpha(Jm+h)}
\end{align}
are the intensive transition rates. The stationary states of the systems correspond to stable fixed points of the mean field dynamics. The fixed points $m^*$ (stable or unstable) are given by the condition $\dot{m}=0$, and they are stable when
 \begin{align}
\left(  \frac{\partial \dot{m}}{\partial m} \right)_{m=m^*} <0 \,.
 \end{align}
The latter condition qualitatively means that the magnetization tends to return to the fixed point after a small perturbation from it.

Due to its nonlinear nature, the mean field equation~\eqref{eq:meanfield} may admit multiple stable fixed points; unstable fixed points then define the boundaries between basins of attractions of the stable fixed points. This contrasts with the master equation description of the model, which predicts a unique stationary state of the system. This apparent incongruity is referred to as the Keizer's paradox~\cite{keizer1978thermodynamics}, and is related to the noncommutativity of two limits: the long-time limit $t \rightarrow \infty$ and the thermodynamic limit $N \rightarrow \infty$~\cite{herpich2020njp,FalascoReview}. The mean field approach corresponds to taking the thermodynamic limit before the long time limit. Then the dynamics may indeed become nonergodic, and thus the stationary state may not be unique.

The approach that characterizes the macroscopic behavior of the system while respecting the uniqueness of the stationary state is provided by large deviation theory~\cite{Touchette2009,FalascoReview,hanggi1982stochastic,gang1987stationary,ge2009thermodynamic}. It enables one to take the long time limit $t \rightarrow \infty$ before applying the large $N$ limit. As this method has been thoroughly discussed in the literature cited above, here we focus only on the most relevant aspects. Within the large deviation framework, the asymptotic scaling of stationary magnetization probabilities is described by the formula
\begin{align} \label{eq:largedev}
    p_M^\text{st} \propto e^{-N V(m)} \,,
\end{align}
where $V(m)$ is the nonequilibrium quasipotential. Since for large $N$ the probability distribution is narrowly focused around the minimum of $V(m)$, the stationary value of normalized magnetization $m_0$ corresponds to the global minimum of the nonequilibrium quasipotential $V(m)$:
\begin{align}
m_0=\argmin_{m \in [-1,1]} V(m) \,.
\end{align}
The other local minima of $V(m)$ correspond to metastable states of the system, whose lifetimes increase exponentially with $N$~\cite{hanggi1984bistable, hinch2005exponentially,kurchan2009equilibrium}. We further note that all minima (maxima) of the quasipotential exactly correspond to stable (unstable) fixed points of mean field equations~\cite{hanggi1982stochastic,FalascoReview}.

The quasipotential can be obtained by noting that the stationary probabilities obey the detailed balance condition $W_{M,M+2}p^\text{st}_{M+2}  =W_{M+2,M} p^\text{st}_M $. Thus, they can be expressed as~\cite{ge2009thermodynamic, landauer1962fluctuations,hanggi1982stochastic}
\begin{align} \nonumber
&p_M^\text{st} = p_{-N}^\text{st} \frac{W_{-N+2,-N} W_{-N+4,-N+2} \ldots W_{M,M-2}}{W_{-N,-N+2} W_{-N+2,-N+4} \ldots W_{M-2,M}} \\ \nonumber
&=p_{-N}^\text{st} \exp \left(\ln \frac{W_{-N+2,-N}}{W_{-N,-N+2}} + \ldots + \ln\frac{ W_{M,M-2}}{ W_{M-2,M}} \right) \,.
\end{align}
In the large $N$ limit, one can convert the above Riemann sum in the parenthesis into a definite integral. Taking into account Eq.~\eqref{eq:largedev}, this yields
\begin{align}
V(m)=\frac{1}{2} \int_{-1}^m dq \ln \frac{w_-(q)}{w_+(q)} \,,
\end{align}
with the intensive transition rates $w_\pm(m)$ defined below Eq.~\eqref{eq:meanfield}. The factor $1/2$ before the integral results from the fact that each jump changes magnetization by $\pm 2$, and thus the above Riemann sum contains $ (1+m)N/2$ elements. The quasipotential is further normalized by adding a constant so that $V(m_0)=0$.

The large deviation approach can be further used to describe nonequilibrium thermodynamics of the system~\cite{FalascoReview}. In particular, using the master equation, the stationary heat current from the bath $\alpha$ can be calculated as
\begin{align}
\langle \dot{Q}_\alpha \rangle=\sum_{M} \sum_{\pm} p_M^\text{st} W^\alpha_{M \pm 2,M} (E_{M \pm 2}-E_M) \,.
\end{align}
The normalized stationary heat current $\langle \dot{q}_\alpha \rangle \equiv \langle \dot{Q}_\alpha \rangle/N$ can then be calculated in the limit $N \rightarrow \infty$ by converting the above sum into an integral and using Eq.~\eqref{eq:largedev}. This yields
\begin{align} \label{eq:meanfhcur} \nonumber
&\langle \dot{q}_\alpha \rangle  \\ \nonumber &=-2\lim_{N \rightarrow \infty} \int_{-1}^{1} e^{-NV(m)}(Jm+h) \left[ w_+^\alpha(m)-w_-^\alpha(m) \right] dm \\
&=-2(Jm_\text{0}+h) \left[ w_+^\alpha(m_0)-w_-^\alpha(m_0) \right] \,,
\end{align}
where the integral is calculated using the Laplace method.

 \subsection{Nonequilibrium phase transitions} \label{sec:phasetr}
To make this paper self-contained, we now discuss the behavior of the model at nonequilibrium phase transitions, which has been previously analyzed in Refs.~\cite{garrido1987stationary,blote1990critical,tamayo1994twotemperature,garrido1996complexity,vroylandt2018collective,vroylandt2020efficiency,aron2020landau, herpich2020njp,beyen2024phase}. At zero magnetic field ($h=0$) the system admits two phases, the paramagnetic phase with $m_0=0$ and the ferromagnetic phase with $m_0 \neq 0$. The system is in the paramagnetic phase if the quasipotential $V(m)$ has a minimum at $m=0$. This yields the condition of stability of the paramagnetic phase:
\begin{align}
\left[\frac{\partial^2 V(m)}{\partial m^2} \right]_{m=0} >0 \,.
\end{align}
One finds that the paramagnetic phase is stable for $\beta_\text{eff}<\beta_c$ ($T_\text{eff}>T_c$), where
\begin{align}
\beta_\text{eff}=\frac{\sum_{\alpha=1}^d \Gamma_\alpha \beta_\alpha}{\sum_{\alpha=1}^d \Gamma_\alpha}
\end{align}
 is the effective inverse temperature, and the critical inverse temperature
 \begin{align}
\beta_c=J^{-1}
 \end{align}
is unchanged with respect to the equilibrium model. This means that the effective inverse temperature $\beta_\text{eff}$ is the weighted average of the inverse temperatures of different reservoirs, $\beta_\alpha$, the weights being proportional to the couplings to the respective reservoirs. In equilibrium, with $\beta_\alpha=\beta$, the effective temperature is equal to the equilibrium temperature ($\beta_\text{eff}=\beta$) independent of the kinetic rates $\Gamma_\alpha$. We also notice that out of equilibrium, for certain values of $\Gamma_\alpha$ and $\beta_\alpha$, the stable paramagnetic fixed point may coexist with the ferromagnetic phase, and thus the transition between them may become discontinuous~\cite{garrido1996complexity}. Our study focuses on the parameter regime where the paramagnetic-ferromagnetic phase transition is continuous.

Let us now focus on the case of two baths ($d=2$) and take the symmetric coupling $\Gamma_1=\Gamma_2=\Gamma$. We further parameterize the bath inverse temperatures as $\beta_1=\beta_\text{eff}-\Delta \beta/2$ and $\beta_2=\beta_\text{eff}+\Delta \beta/2$. Interestingly, in this case the ratio
\begin{align}
\frac{w_-(m)}{w_+(m)}=\ \frac{1+m}{1-m} e^{-2 \beta_\text{eff} (Jm+h)} \,,
\end{align}
and thus the whole quasipotential $V(m)$, is the same as for the equilibrium model with $\beta=\beta_\text{eff}$. It is given (up to a constant) by the formula
\begin{align}
V(m)=&-\beta_\text{eff} \left(\frac{J}{2}m^2+hm \right) \\ \nonumber &+\frac{1+m}{2} \ln \frac{1+m}{2}+\frac{1-m}{2} \ln \frac{1-m}{2} \,.
\end{align}
Here, the first term corresponds to the scaled energy of the system $\lim_{N \rightarrow \infty} \beta_\text{eff} E_M/N$, with $E_M$ given by Eq.~\eqref{eq:modelenergy}. The next two terms correspond to scaled Boltzmann entropy with a minus sign, $-\lim_{N \rightarrow \infty}(\ln \Omega_M)/N$, with $\Omega_M$ given by Eq.~\eqref{eq:omegam}. Thus, in equilibrium with $\beta_\text{eff}=\beta$, the quasipotential corresponds to the scaled free energy, $V(m)=\lim_{N \rightarrow \infty} (\beta F_M/N)$, where $F_M=E_M-\beta^{-1} \ln \Omega_M$. Consequently, the stationary value of magnetization and its critical behavior is the same as for the equilibrium model with $\beta=\beta_\text{eff}$, where minimization of the quasipotential is equivalent to minimization of the free energy.

%%%%%%%%%%%%%%%%%%%%%%%%%%%%%%%%%%%%%%%%%%%%%%%%%%%%%%%%%%%%%%%%%%%%
\begin{figure}
	\centering
	\includegraphics[width=0.88\linewidth]{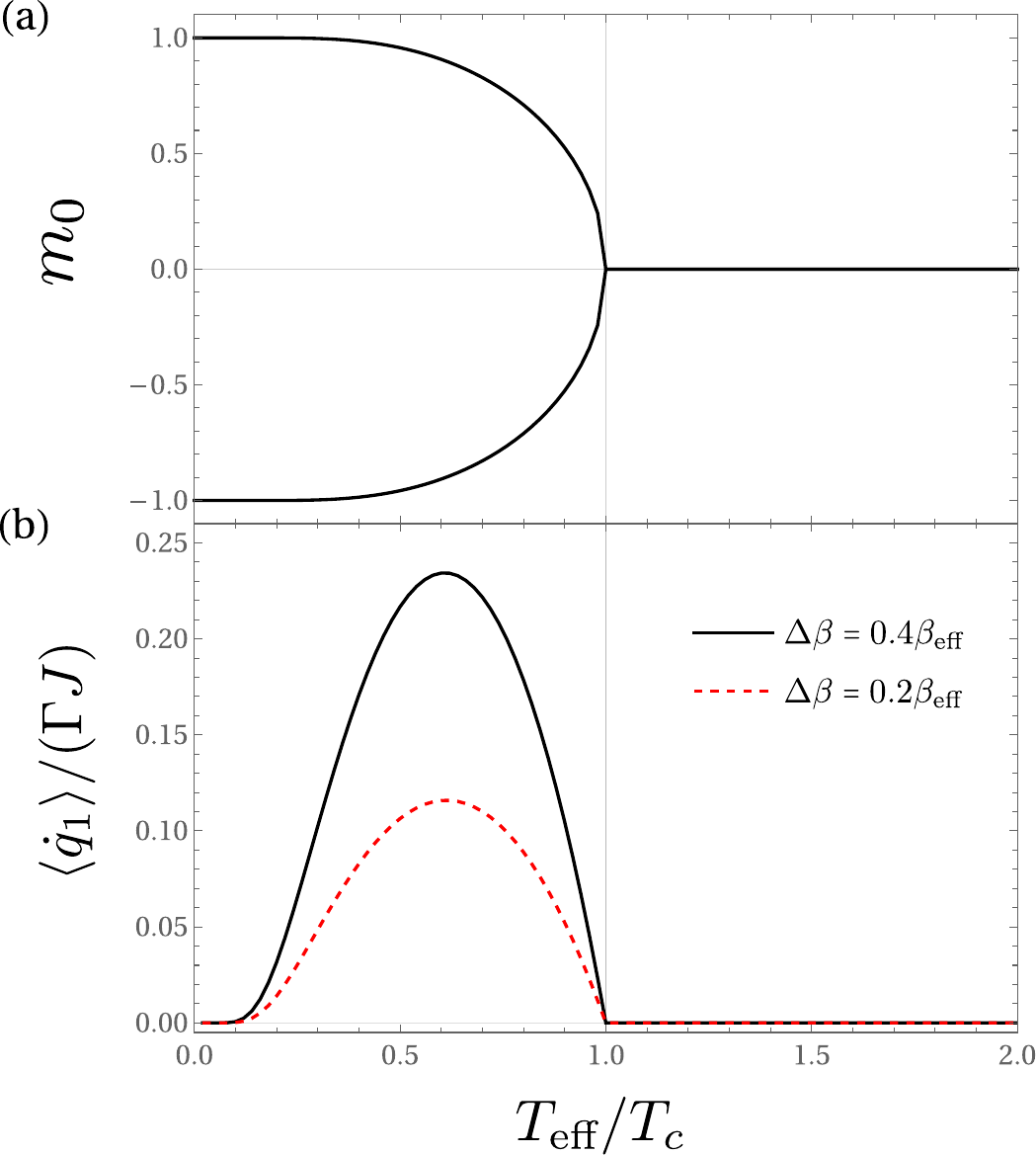}
	\caption{The normalized stationary magnetization $m_0$ (a) and the heat current $\langle \dot{q}_1 \rangle=-\langle \dot{q}_2 \rangle$ (b) as a function of the effective temperature $T_\text{eff}$ for $h=0$ and $\Gamma_1=\Gamma_2=\Gamma$.}
	\label{fig:phasetr-temp}
\end{figure}
%%%%%%%%%%%%%%%%%%%%%%%%%%%%%%%%%%%%%%%%%%%%%%%%%%%%%%%%%%%%%%%%%%%%
In Fig.~\ref{fig:phasetr-temp} we present stationary magnetization $m_0$ and heat current $\langle \dot{q}_1 \rangle=-\langle \dot{q}_2 \rangle$ as a function of the effective temperature $T_\text{eff}$ for $h=0$. As can be observed, both quantities exhibit a continuous phase transition from 0 to a finite value at $T_\text{eff}=T_c$. We show two branches of $m_0 = \pm |m_0|$ that correspond to degenerate minima of $V(m)$. The presence of such degenerate minima is characteristic for continuous symmetry-breaking phase transitions. In contrast, the heat current is single-valued, as it is invariant to the simultaneous reversal of the magnetic field and magnetization ($h \rightarrow -h$ and $M \rightarrow -M$). Intuitively, the magnitude of the heat current increases with the temperature difference $\Delta \beta$, and it vanishes at equilibrium ($\Delta \beta=0)$.

%%%%%%%%%%%%%%%%%%%%%%%%%%%%%%%%%%%%%%%%%%%%%%%%%%%%%%%%%%%%%%%%%%%%
\begin{figure}
	\centering
	\includegraphics[width=0.88\linewidth]{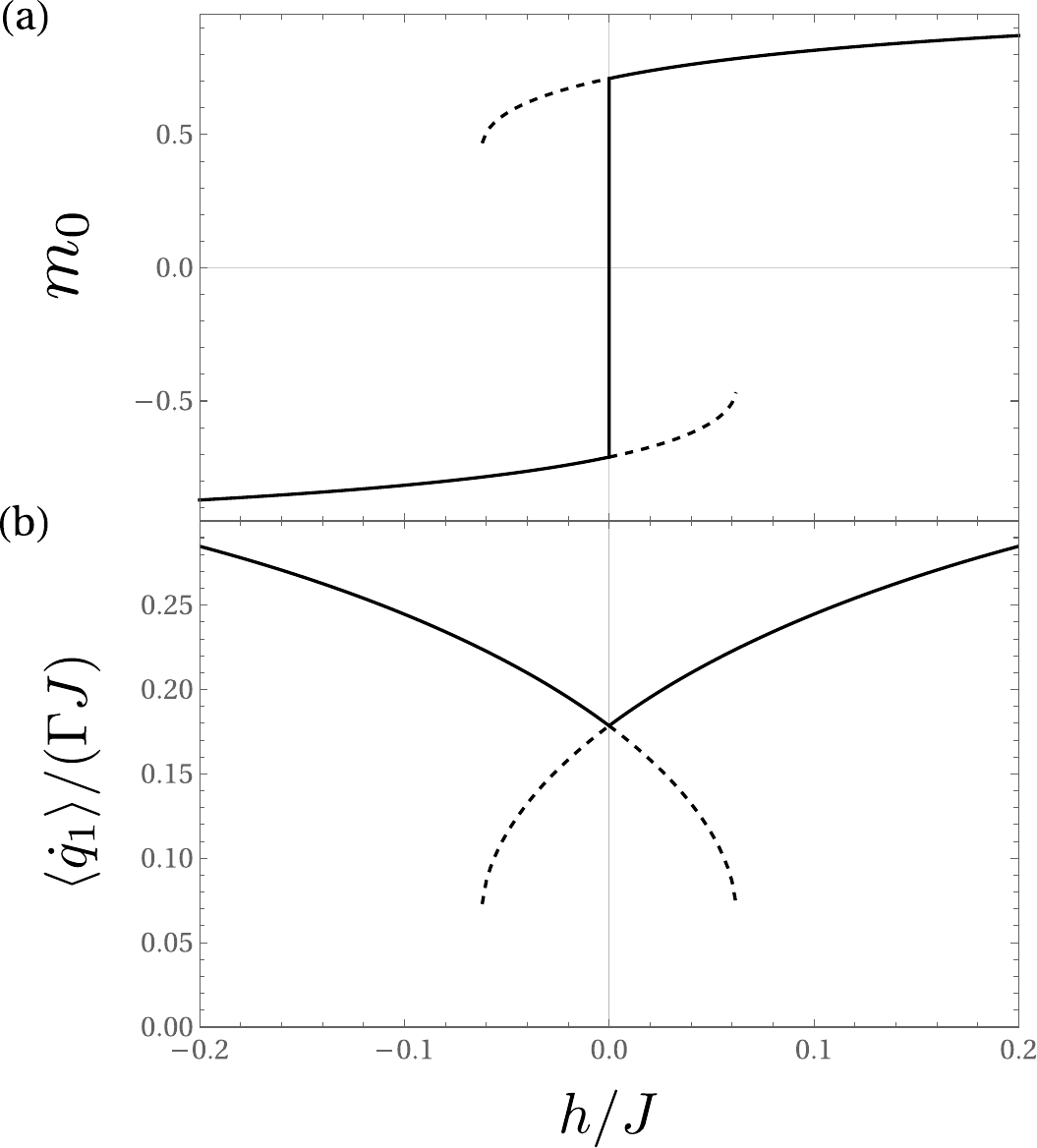}
	\caption{The normalized stationary magnetization $m_0$ (a) and the heat current $\langle \dot{q}_1 \rangle=-\langle \dot{q}_2 \rangle$ (b) as a function of the magnetic field $h$ for $T=0.8 T_\text{eff}$, $\Delta \beta=0.4 \beta_\text{eff}$ and $\Gamma_1=\Gamma_2=\Gamma$. The dashed lines represent the metastable solutions. The dashed line in (b) is calculated using Eq.~\eqref{eq:meanfhcur}, but with $m_0$ replaced by the metastable value of magnetization.}
	\label{fig:phasetr-field}
\end{figure}
%%%%%%%%%%%%%%%%%%%%%%%%%%%%%%%%%%%%%%%%%%%%%%%%%%%%%%%%%%%%%%%%%%%%
The other phase transition takes place when the magnetic field $h$ is changed for a constant effective temperature $T_\text{eff}<T_c$. This is presented in Fig.~\ref{fig:phasetr-field}. Interestingly, this phase transition can be regarded as either discontinuous or continuous, depending on the observable that is considered. As in equilibrium, the stationary magnetization exhibits a discontinuous jump at $h=0$. We can also observe the presence of metastable solutions [that is, the local minima of $V(m)$], represented by dashed lines, that correspond to the analytic continuations of the stationary solutions. In contrast, the heat current is continuous but nonanalytic at $h=0$, having a kink at this point. This is a consequence of the aforementioned symmetry with respect to the magnetization reversal. A similar kink of entropy production has been previously observed in other nonequilibrium phase transition models~\cite{chrochik2005entropy, noa2019entropy}. 

\section{Fluctuations and responses: methods} \label{sec:fluctmeth}

\subsection{Spectral approach} \label{subsec:spec}
Let us now discuss the methods used to characterize the heat current fluctuations, which are the main focus on the paper, as well as the current responses, which will be useful at some points. We first focus on the spectral approach~\cite{esposito2007fluctuation,flindt2008counting,walldorf2020noise}, which is applicable for finite system sizes $N$. This method has recently been thoroughly reviewed in Ref.~\cite{landi2023current}; here, we just briefly present the most relevant results.

Denote by $Q_\alpha(t)$ the stochastic (fluctuating) value of the heat transferred from the bath $\alpha$ to the system in the time interval $[0,t]$. To describe the heat current fluctuations, we define the scaled cumulant generating function of $Q_\alpha(t)$,
\begin{align} \label{eq:scaledcumgen}
	S_\alpha(\chi_\alpha) \equiv \lim_{t \rightarrow \infty} \frac{1}{t} \ln \int_{-\infty}^\infty \rho[Q_\alpha(t)] e^{\chi_\alpha Q_\alpha(t)} d Q_\alpha(t) \,,
\end{align}
where $\rho[Q_\alpha(t)]$ is the probability density of $Q_\alpha(t)$ and $\chi_\alpha$ is the counting field. This function can be expanded as a power series
\begin{align} \label{eq:scaledcumgenexp}
S_\alpha(\chi_\alpha)=\sum_{n=1}^\infty \frac{\llangle Q_\alpha^n \rrangle \chi_\alpha^n}{n!} \,,
\end{align}
where the coefficients $\llangle Q_\alpha^n \rrangle$ are the scaled cumulants of heat current from the bath $\alpha$. In particular, the first two scaled cumulants correspond to the average heat current $\langle \dot{Q}_\alpha \rangle$ and the current variance $\Var(Q_\alpha)$, respectively:
\begin{align}
\llangle Q^1_\alpha \rrangle & = \langle \dot{Q}_\alpha \rangle \equiv \lim_{t \rightarrow \infty} \frac{\langle Q_\alpha(t) \rangle}{t} \,, \\
\llangle Q_\alpha^2 \rrangle & = \Var(Q_\alpha) \equiv \lim_{t \rightarrow \infty} \frac{\left \langle \left[ Q_\alpha(t) - \langle Q_\alpha(t) \rangle \right]^2 \right \rangle}{t} \,.
\end{align}

Let us now define the counting-field-dependent rate matrix $\mathbb{W}(\boldsymbol{\chi})$, where $\boldsymbol{\chi}=(\chi_1,\ldots,\chi_d)$ is the vector of the counting fields. Its elements read as
\begin{align}
	\begin{cases}
		[\mathbb{W}(\boldsymbol{\chi})]_{kl} &=\sum_{\alpha=1}^d W^\alpha_{kl} e^{\chi_\alpha (E_k-E_l)} \quad \text{for} \quad k \neq l \,, \\
		[\mathbb{W}(\boldsymbol{\chi})]_{kk} &=-\sum_{l \neq k} W_{lk} \,,
	\end{cases}
\end{align}
where $W_{kl}$, $W_{kl}^\alpha$ and $E_k$ are the transition rates and the state energies defined in Sec.~\ref{subsec:mod}, and $k,{ }l \in \{-N,-N+2,\ldots,N\}$ denote the magnetization values. The rate matrix $\mathbb{W}(\boldsymbol{\chi}=\boldsymbol{0})$, where $\boldsymbol{0} \equiv (0,\ldots,0)$, is simply denoted as $\mathbb{W}$. We further define the left and right eigenvectors of $\mathbb{W}$ as
\begin{align}
	\mathbb{W}|x_j \rrangle =\lambda_j |x_j \rrangle, \quad \llangle y_j |\mathbb{W} =\lambda_j \llangle y_j | \,.
\end{align}
The eigenvectors are normalized as $\llangle y_j|x_j \rrangle=1$. The eigenvalues of the rate matrix are here ordered in the decreasing order of their real parts: $\re(\lambda_0) \geq \re(\lambda_1) \geq \ldots \geq \re(\lambda_N)$. The dominant eigenvalue $\lambda_0$ is equal to 0 by virtue of the Perron–Frobenius theorem. The corresponding right eigenvector $|x_0 \rrangle$ is the stationary state of the system $|p_\text{st} \rrangle \equiv (p_{-N}^\text{st},\ldots,p_N^\text{st})^\intercal$. The associated left eigenvector reads $\llangle y_0|=(1,\ldots,1)$, and is further denoted as $\llangle \mathds{1}|$.

The scaled cumulants are given by the recursive relation\footnote{Strictly speaking, the formula below is applicable when the rate matrix is diagonalizable, and thus its right and left eigenvectors form a complete basis. The opposite situation, when the rate matrix becomes defective and some eigenvectors become degenerate, may occur for certain points in the parameter space, called the exceptional points~\cite{minganti2019quantum}. However, the master equation considered here corresponds to the one-dimensional random walk, for which the rate matrix is always diagonalizable~\cite{ledermann1954spectral}.}
\begin{align}
\llangle Q_\alpha^n \rrangle=\sum_{m=1}^n {n \choose m} \llangle \mathds{1}| \mathbb{W}^{(m)}_\alpha|p_\text{st}^{(n-m)} \rrangle \,,
\end{align}
where $|p_\text{st}^{(0)} \rrangle \equiv |p_\text{st} \rrangle$ and
\begin{align} \label{eq:pnrecur}
|p_\text{st}^{(n)} \rrangle &\equiv \mathbb{W}^D \sum_{m=1}^n {n \choose m} \left(\llangle Q_\alpha^m \rrangle-\mathbb{W}^{(m)}_\alpha \right)|p_\text{st}^{(n-m)} \rrangle \,, \\
\mathbb{W}_\alpha^{(n)} & \equiv \left[\frac{\partial^n}{\partial \chi_\alpha^n} \mathbb{W}(\boldsymbol{\chi}) \right]_{\boldsymbol{\chi}=\boldsymbol{0}} \,,
\end{align}
where $\mathbb{W}^D=\sum_{j=1}^N \lambda_j^{-1} |x_j \rrangle \llangle y_j|$ is the Drazin inverse of $\mathbb{W}$ (see Refs.~\cite{landi2023current, crook2018drazin} for a discussion of its properties). In particular, the average heat current and the current variance read as
\begin{align}	\label{eq:avcur}
	\langle \dot{Q}_\alpha \rangle &= \llangle \mathds{1}| \mathbb{W}^{(1)}_\alpha|p_\text{st} \rrangle \,,\\ \nonumber \label{varcur}
	\Var(Q_\alpha) &=\llangle \mathds{1}| \mathbb{W}_\alpha^{(2)}|p_\text{st} \rrangle \\ &-2 \sum_{j \neq 0} \frac{\llangle \mathds{1}| \mathbb{W}^{(1)}_\alpha|x_j \rrangle \llangle y_j |\mathbb{W}^{(1)}_\alpha|p_\text{st} \rrangle}{\lambda_j} \,.
\end{align}

Finally, since the system is extensive, we further define the normalized scaled cumulants for a single spin $\llangle q_\alpha^n \rrangle \equiv \llangle Q_\alpha^n \rrangle/N$. In particular, the normalized heat current and the current variance are denoted as $\langle \dot{q}_\alpha \rangle \equiv \langle \dot{Q}_\alpha \rangle/N$ and $\Var(q_\alpha) \equiv \Var(Q_\alpha)/N$.

\subsection{Current responses} \label{subsec:curresp}
A similar approach can be used to calculate the current responses, that is, their derivatives over some control parameter $\zeta$ (in our case, it will be the temperature difference). Using Eq.~\eqref{eq:avcur}, they can be calculated as
\begin{align}
\frac{\partial^n \langle \dot{Q}_\alpha \rangle}{\partial \zeta^n}=\llangle \mathds{1}|\sum_{m=0}^n {n \choose m} \frac{\partial^{m} \mathbb{W}^{(1)}_\alpha}{\partial \zeta^{m}} \frac{\partial^{(n-m)} |p_\text{st} \rrangle}{\partial \zeta^{(n-m)}} \,.
\end{align}
The derivatives of the stationary state vectors $|p_\text{st} \rrangle$ can be determined using the stationarity condition
\begin{align}
\frac{\partial^n}{\partial \zeta^n} \mathbb{W}  |p_\text{st} \rrangle =\sum_{m=0}^n {n \choose m} \frac{\partial^{m} \mathbb{W}}{\partial \zeta^{m}} \frac{\partial^{(n-m)} |p_\text{st} \rrangle}{\partial \zeta^{(n-m)}}=0 \,.
\end{align}
We now note that the derivatives of the stationary state $(\partial^n |p_\text{st} \rrangle)/(\partial \zeta^n)$ (for $n>0$) are traceless vectors due to probability conservation, and thus $\llangle \mathds{1}|(\partial^n |p_\text{st} \rrangle)/(\partial \zeta^n)=0$. Using the identity $\mathbb{W}^D\mathbb{W}=\mathds{1}-|p_\text{st} \rrangle \llangle \mathds{1}|$~\cite{landi2023current,crook2018drazin}, where $\mathds{1}$ is the identity matrix, we see that $\mathbb{W}^D\mathbb{W} (\partial^n |p_\text{st} \rrangle)/(\partial \zeta^n)=(\partial^n |p_\text{st} \rrangle)/(\partial \zeta^n)$. Applying $\mathbb{W}^D$ from the left to the second term of the equation above, we obtain a recursive formula similar to Eq.~\eqref{eq:pnrecur},
\begin{align}
\frac{\partial^{n} |p_\text{st} \rrangle}{\partial \zeta^{n}}=-\mathbb{W}^D \sum_{m=1}^n {n \choose m} \frac{\partial^{m} \mathbb{W}}{\partial \zeta^{m}} \frac{\partial^{(n-m)} |p_\text{st} \rrangle}{\partial \zeta^{(n-m)}} \,.
\end{align}
We note that an alternative method to calculate the linear response of the stationary state vector has been recently presented in Refs.~\cite{aslyamov2024nonequilibrium,aslyamov2024general}.

\subsection{Equilibrium fluctuation-response relation}
Let us now discuss alternative methods to characterize the heat current fluctuations, which will be useful for large system size. First, at thermodynamic equilibrium, the heat current fluctuations are exactly related to the linear response to the temperature gradients through the symmetry of the scaled cumulant generating function~\cite{saryal2019thermodynamic,saito2008symmetry}. For a system connected to two baths $\alpha \in \{1,2\}$ this fluctuation-response relation reads 
\begin{align}
&\Var(Q_1)=\Var(Q_2) =2\left(\frac{\partial \langle \dot{Q}_1 \rangle}{\partial \Delta \beta} \right)_{\Delta \beta=0} \,,
\end{align}
where $\Delta \beta=\beta_2-\beta_1$. We emphasize that the above formula is valid regardless of system size. In particular, it enables characterizing current fluctuations in the thermodynamic limit, where heat currents and their responses can be determined using the macroscopic formula~\eqref{eq:meanfhcur}.

A particularly simple solution can be found in the case of symmetric coupling to the baths ($\Gamma_1=\Gamma_2)$. Then, as discussed in Sec.~\ref{sec:phasetr}, in the thermodynamic limit $N \rightarrow \infty$, the stationary magnetization $m_0$ does not depend on the temperature difference $\Delta \beta$. Using Eq.~\eqref{eq:meanfhcur}, the equilibrium heat current fluctuations are then given by a compact formula
\begin{align} \label{eq:eqfluctsym}
&\Var(q_\alpha)=2(Jm_0+h)^2 \left[w^\alpha_+(m_0)+w^\alpha_-(m_0) \right] \,,
\end{align}
where only the equilibrium magnetization $m_0$ needs to be determined numerically.

\subsection{Path integral approach} \label{subsec:pathint}
Another method used is the recently developed approach to current fluctuations~\cite{herpich2020njp,lazarescu2019large,vroylandt2020efficiency,gopal2022large} based on Martin--Siggia--Rose path integral formulation of stochastic dynamics~\cite{martin1973statistical,peliti1985path,weber2017master}. Since details of this approach were thoroughly described in the above references, we focus only on the most relevant aspects. Analogously to the mean field approach described in Sec.~\ref{subsec:mfldev}, this method applies the thermodynamic limit $N \rightarrow \infty$ before the long time limit $t \rightarrow \infty$, and thus describes the current fluctuations around the deterministic trajectory corresponding to the solution of the mean field equation~\eqref{eq:meanfield}. Consequently, it characterizes \textit{conditional fluctuations} for a system initialized in a basin of attraction of a given fixed point and measured during an observation time that is short compared to the lifetime of this fixed point, but long compared with the other time scales of the system; such conditional fluctuations have been recently analyzed in Refs.~\cite{vroylandt2020efficiency,fiore2021current}. 

Within the path integral approach, the heat current fluctuations are characterized by the normalized scaled cumulant generating function
\begin{align} \label{eq:cumgenpathdef}
s_\alpha(\chi_\alpha) \equiv \lim_{N \rightarrow \infty} \frac{1}{N} S_\alpha(\chi_\alpha) \,,
\end{align}
where $S_\alpha(\chi_\alpha)$ is the unnormalized scaled cumulant generating function defined in Eq.~\eqref{eq:scaledcumgen}. The normalized current cumulants can be then calculated using Eq.~\eqref{eq:scaledcumgenexp} as
\begin{align} \label{eq:cumpath}
    \llangle q^n_\alpha \rrangle= \left[\frac{\partial^n}{\partial \chi_\alpha^n} s_\alpha(\chi_\alpha) \right]_{\chi_\alpha=0} \,.
\end{align}
To determine $s_\alpha(\chi_\alpha)$, one employs the counting field-dependent (biased) Freidlin-Wentzell Hamiltonian~\cite{herpich2020njp}
\begin{align}
H_{\chi_\alpha}(m,p)= \sum_{\nu=1}^{d} \sum_{\pm} w_\pm^\nu (m) \left[e^{\pm 2 p} e^{\pm \delta_{\nu \alpha} \chi_\alpha \mathcal{E}(m)} -1\right] \,,
\end{align}
where $\delta_{\nu \alpha}$ is the Kronecker delta and
\begin{align}
\mathcal{E}(m)=\lim_{N \rightarrow \infty} (E_{M+2}-E_M)=-2(Jm+h)
\end{align}
is the energy change due to flipping one spin up (i.e., transition $M \rightarrow M+2$). The function $s_\alpha(\chi_\alpha)$ is then calculated as
\begin{align} \label{eq:cumgenpath}
s_\alpha(\chi_\alpha) =H_{\chi_\alpha}(m_{\chi_\alpha}^*,p_{\chi_\alpha}^*) \,,
\end{align}
where $\{m_{\chi_\alpha}^*,p_{\chi_\alpha}^* \}$ is the fixed point of the equations of motions
\begin{align} \label{eq:eqmotbiased}
\dot{m} =\frac{\partial}{\partial p} H_{\chi_\alpha}(m,p) \,, \quad
\dot{p} =-\frac{\partial}{\partial m} H_{\chi_\alpha}(m,p) \,.
\end{align}
We notice that, analogously to the mean field equation~\eqref{eq:meanfield}, the above equations can have multiple fixed points. Since the current cumulants are determined by the behavior of $s_\alpha(\chi_\alpha)$ close to $\chi_\alpha=0$ [see Eq.~\eqref{eq:cumpath}], to determine the steady-state fluctuations one considers the fixed point lying close the fixed point of the unbiased dynamics [Eq.~\eqref{eq:eqmotbiased} with $\chi_\alpha=0$] that corresponds to the stationary state of the system. The latter fixed point has the coordinates $m_{\chi_\alpha=0}=m_0$ and $p_{\chi_\alpha=0}=0$. The finding of the fixed point can be further simplified using the analytic solution for $p_{\chi_\alpha}^*$,
\begin{align}
p_{\chi_\alpha}^*=\frac{1}{4} \ln \frac{\sum_{\nu=1}^{d} w_-^\nu (m_{\chi_\alpha}) e^{- \delta_{\nu \alpha} \chi_\alpha \mathcal{E}(m_{ \chi_\alpha})}}{\sum_{\nu=1}^{d} w_+^\nu (m_{\chi_\alpha}) e^{\delta_{\nu \alpha} \chi_\alpha \mathcal{E}(m_{\chi_\alpha})}} \,.
\end{align}
The solution $m_{\chi_\alpha}^*$ can then be found numerically. Consequently, one can determine $s_\alpha(\chi_\alpha)$ numerically in the vicinity of $\chi_\alpha=0$ and then numerically estimate the derivatives in Eq.~\eqref{eq:cumpath} using the finite difference method. In particular, to the lowest order of precision, the normalized current variance can be estimated as
\begin{align} \label{eq:findif1st}
\Var(q_\alpha) = \frac{s_\alpha(0)-2s_\alpha(\epsilon)+s_\alpha(2\epsilon)}{\epsilon^2}+O(\epsilon) \,,
\end{align}
with a small parameter $\epsilon$. The higher-order finite difference methods are easily accessible in the literature.

\subsection{Two-state model} \label{subsec:tsm}
As a last method, let us present an effective two-state model proposed in Ref.~\cite{nguyen2020exponential}. This model is applicable in the regime where the system has two stable fixed points, i.e., the quasipotential $V(m)$ has two minima. We further denote these minima as $m_+$ and $m_-$, and the maximum of $V(m)$ that separates them as $m_*$. The model assumes that, for large system sizes $N$, the dominant contribution to noise results from stochastic switching between the heat current values associated with the fixed points $m_+$ and $m_-$. This switching is described as a Markov jump process among two discrete states corresponding to different fixed points. The jumps from the state $m_\pm$ to $m_\mp$ occur with the transition rate $r_\pm$. For large $N$, these transition rates can be approximated as~\cite{hinch2005exponentially,hanggi1984bistable,nguyen2020exponential}\footnote{We notice that our Eq.~\eqref{eq:ratestsm} appears to be larger than the corresponding Eq.~(37) in Ref.~\cite{nguyen2020exponential} by a prefactor 4. The reason is that Ref.~\cite{nguyen2020exponential} considered the case where the difference between successive values of the number of molecules $X$ was equal to 1, while in our case the difference between successive values of magnetization $M$ equals $2$. Consequently, Eq.~\eqref{eq:ratestsm} can be obtained from Eq.~(37) in Ref.~\cite{nguyen2020exponential} by using parameterization $X=M/2$, $x=m/2$, so that $V''(x)=4V''(m)$.}
\begin{align} \label{eq:ratestsm}
r_\pm \approx \frac{2e^{-N [V(m_*)-V(m_\pm)]} w_+(m_\pm) \sqrt{-V''(m_\pm)V''(m_*)}}{\pi N} \,.
\end{align}
This expression implies that transition rates are suppressed exponentially with $N$ multiplied by the quasipotential barrier; the latter corresponds to the difference between the value of $V(m)$ at the fixed point $m_\pm$ from which the jump occurs and the maximum $m_*$. Each fixed point is further assumed to be associated with a well-defined heat current value
\begin{align}
\langle \dot{q}_\alpha \rangle_\pm=-2(J m_\pm+h)[w_+^\alpha(m_\pm)-w_-^\alpha(m_\pm)] \,,
\end{align}
which is the average heat current calculated using Eq.~\eqref{eq:meanfhcur} for a given fixed point $m_\pm$. Using these assumptions, the heat current variance associated with the stochastic switching between the current values $\langle \dot{q}_\alpha \rangle_\pm$ can be calculated as~\cite{nguyen2020exponential}
\begin{align} \label{eq:vartsm}
\Var(q_\alpha) \approx 2 \left(\langle \dot{q}_\alpha \rangle_+-\langle \dot{q}_\alpha \rangle_- \right)^2 \frac{p_+ p_-}{r_++r_-} \,,
\end{align}
where $p_\pm=r_\mp/(r_++r_-)$ are the probabilities of states $m_\pm$.

\section{Results: Fluctuations scaling}\label{sec:fluctscal}
Let us now turn to the main point of the paper, namely, the analysis of heat current fluctuations. For simplicity, in the whole section we focus on the case of two symmetrically coupled baths ($\Gamma_1=\Gamma_2=\Gamma)$. The main quantity analyzed is the normalized current variance $\Var(q_1)$, which is equal to $\Var(q_2)$ due to energy conservation.

\subsection{Temperature-driven transition} \label{subsec:temp}
\begin{figure}
	\centering
	\includegraphics[width=0.9\linewidth]{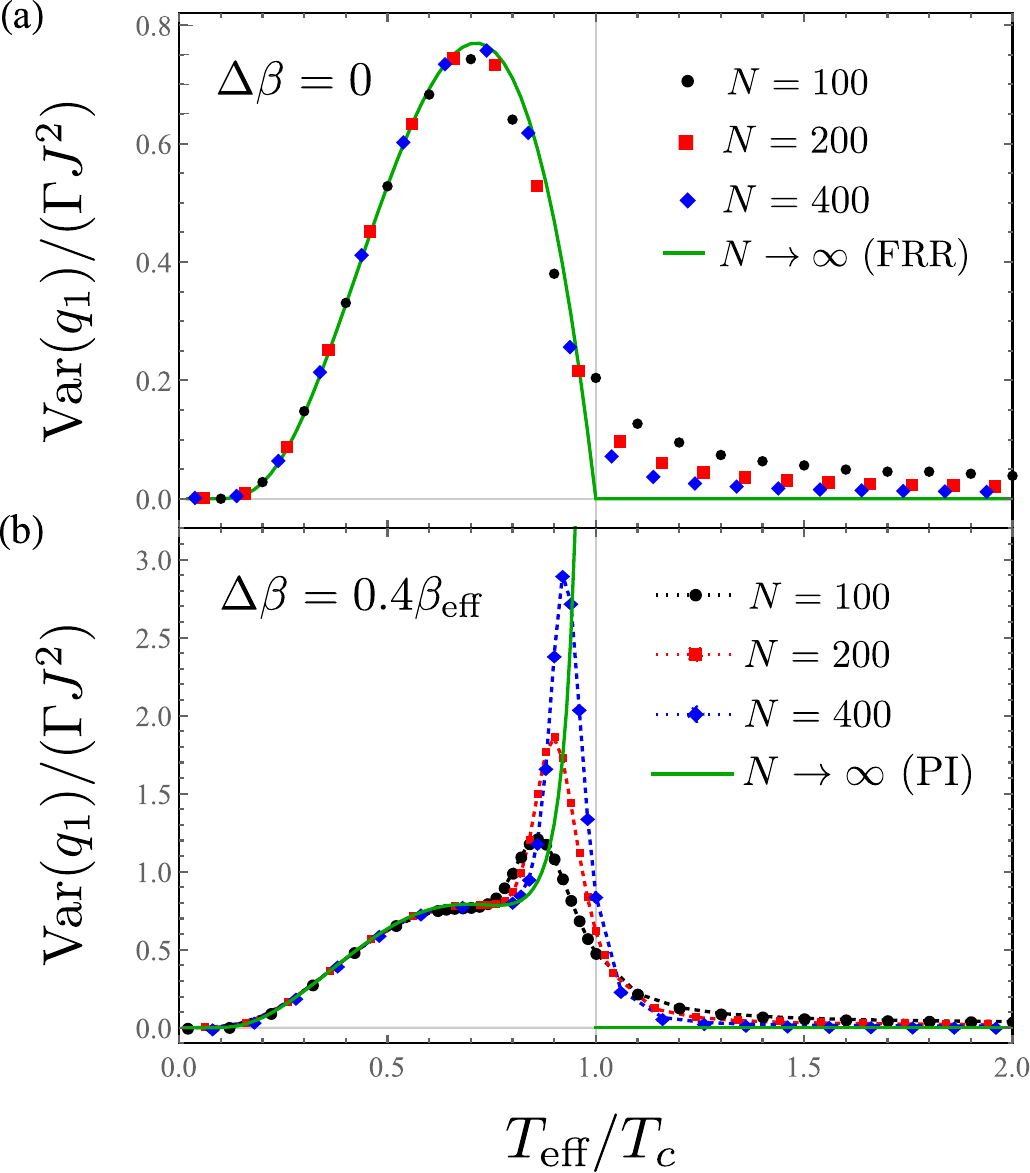}
	\caption{The effective temperature dependence of the heat current variance $\Var(q_1)$ for different system sizes $N$ at equilibrium ($\Delta \beta=0$) (a) and for $\Delta \beta=0.4 \beta_\text{eff}$ (b). The green solid line in (a) represents the equilibrium fluctuation-response relation (FRR), Eq.~\eqref{eq:eqfluctsym}, while in (b) it represents the path integral (PI) results. The dotted lines in (b) are added for eye guidance. Parameters: $h=0$, $\Gamma_1=\Gamma_2$, $\beta_1=\beta_\text{eff}-\Delta \beta/2$, $\beta_2=\beta_\text{eff}+\Delta \beta/2$.}
	\label{fig:tempdep}
\end{figure}

Let us first analyze the dependence of the heat current variance on the effective temperature, evaluated at $h=0$. We consider both the exact results for finite system sizes, provided by the spectral approach from Sec.~\ref{subsec:spec}, and the asymptotic results for the thermodynamic limit $N \rightarrow \infty$. In the equilibrium case, the latter results are evaluated using the fluctuation-response relation, Eq.~\eqref{eq:eqfluctsym}, which was found to fully agree with the path integral approach from Sec.~\ref{subsec:pathint}. In the nonequilibrium case, we use the path integral approach, and the current variance is evaluated using Eq.~\eqref{eq:findif1st} with $\epsilon=0.001$.\footnote{For the values of $\Var(q_1)$ presented in Fig.~\ref{fig:tempdep}, the results are consistent with those for larger and smaller $\epsilon$, as well as with the results obtained using the higher-order finite difference methods. The consistency becomes worse only in a region very close to the criticality, where $\Var(q_1)$ goes beyond the range of Fig.~\ref{fig:tempdep}.} We recall that in the ferromagnetic case the actual steady state corresponds to an equally weighted mixture of two degenerate magnetization states $m_0=\pm |m_0|$. However, the results obtained using the fluctuation-response relation or the path integral approach are the same when evaluated for either of those states; furthermore, the stochastic switching between these states does not contribute to noise for $h=0$, as both states are associated with the same current value (see more elaborate discussion in Sec.~\ref{subsec:field}).

%%%%%%%%%%%%%%%%%%%%%%%%%%%%%%%%%%%%%%%%%%%%%%%%%%%%%%%%%%%%%%%%%%%%
%%%%%%%%%%%%%%%%%%%%%%%%%%%%%%%%%%%%%%%%%%%%%%%%%%%%%%%%%%%%%%%%%%%%	

The results are presented in Fig.~\ref{fig:tempdep} for the equilibrium~(a) and the nonequilibrium~(b) cases.
At equilibrium, the variance tends to saturate at a finite value with increasing system size, and converges asymptotically to the predictions of the fluctuation-response relation for the thermodynamic limit $N \rightarrow \infty$. In particular, in the paramagnetic phase ($T_\text{eff}>T_c$), the normalized variance decreases monotonically with system size, since the current response vanishes in this limit. This results from the fact that, for $N \rightarrow \infty$, the flipping of a spin for $m=0$ does not change the energy of the system. In the ferromagnetic phase $(T_\text{eff} < T_c)$, especially away from the phase transition point, we observe a good quantitative agreement between the results for finite and infinite $N$, even for a relatively low number of spins $N=100$. We further notice that in the thermodynamic limit $N \rightarrow \infty$, the heat current fluctuations exhibit a nonanalytic behavior at $T_\text{eff}=T_c$ (they are equal to zero for $T_\text{eff} \geq T_c$ and finite for $T_\text{eff} <T_c$). This shows that current fluctuations can witness the presence of phase transitions in equilibrium, where there is no average current.

A very different behavior of heat current fluctuations  is observed in the nonequilibrium case [Fig.~\ref{fig:tempdep}~(b)] close to the phase transition point. As one may observe, the current variance evaluated for finite $N$ exhibits a pronounced peak close to the phase transition point, whose magnitude increases with system size. This agrees with the path integral results, which predict the asymptotic divergence of heat current fluctuations when the phase transition point is approached from the ferromagnetic phase. Away from the phase transition point, the results are generally analogous to the equilibrium case: In the paramagnetic phase, the finite-size results converge asymptotically to zero, in agreement with the path integral approach, which predicts the vanishing of heat current fluctuations. In the ferromagnetic phase, one observes a good agreement between the finite-size results and the path integral approach even for relatively small system sizes. Finally, we note that the current variance is discontinuous at the phase transition point. This comes from the fact that the scaled cumulant generating function $s(\chi_1)$ is then nonanalytic at $\chi_1=0$; see Ref.~\cite{herpich2020njp} for more details.
	
\begin{figure}[t]
	\centering
	\includegraphics[width=0.9\linewidth]{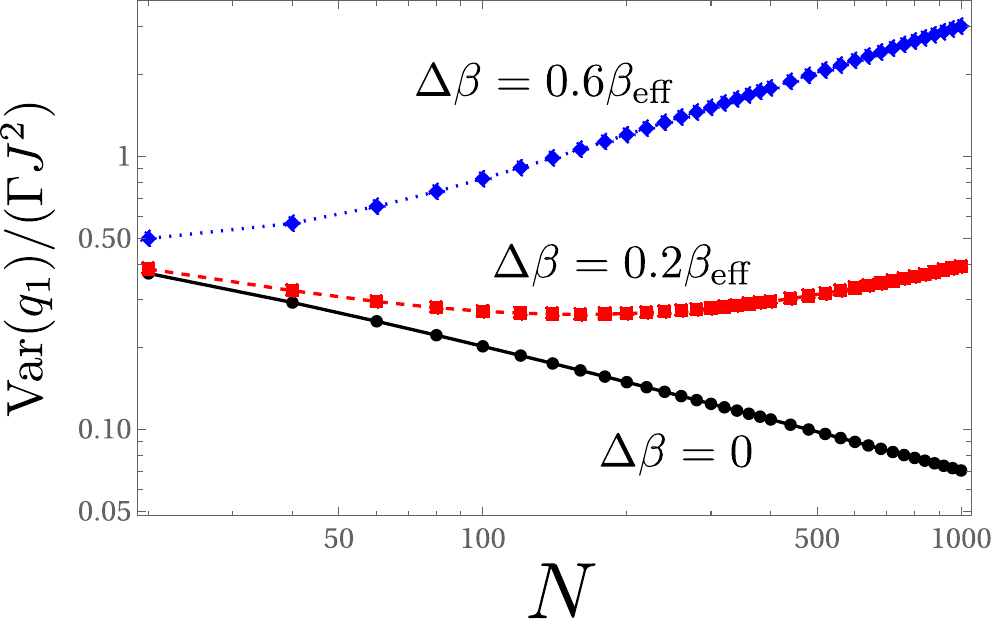}
	\caption{The finite-size scaling (on the log-log scale) of the heat current variance $\Var(q_1)$ at the phase transition point ($T_\text{eff}=T_c$) for different temperature differences $\Delta \beta$.  Other parameters as in Fig.~\ref{fig:tempdep}. Lines added for eye guidance.}
	\label{fig:scaling-var-temp}
\end{figure}

We now focus on the finite-size scaling behavior of the current variance at the phase transition point ($T_\text{eff}=T_c$). This is illustrated in Fig.~\ref{fig:scaling-var-temp}. As shown, in equilibrium, the variance decays as a power law, tending to $0$ for $N \rightarrow \infty$, in agreement with the predictions of the fluctuation-response relation~\eqref{eq:eqfluctsym}. For a large temperature difference $\Delta \beta=0.6 \beta_\text{eff}$ it exhibits, instead, a power-law divergence. Finally, for a small temperature difference $\Delta \beta=0.2 \beta_\text{eff}$ the heat current variance exhibits a nonmonotonic behavior, first decreasing and then increasing with system size.

%%%%%%%%%%%%%%%%%%%%%%%%%%%%%%%%%%%%%%%%%%%%%%%%%%%%%%%%%%%%%%%%%%%%
%%%%%%%%%%%%%%%%%%%%%%%%%%%%%%%%%%%%%%%%%%%%%%%%%%%%%%%%%%%%%%%%%%%%	

To explain this behavior, we note that the heat current variance can be expanded in $\Delta \beta$ as~\cite{saryal2019thermodynamic}
\begin{align} \nonumber
\Var(q_1)=&\Var_\text{eq}(q_1)+  \Delta \beta^2 \left[\frac{\llangle q_1^4 \rrangle_\text{eq}}{12} +\frac{G_3(q_1)}{3}\right] \\ &+O(\Delta \beta^4) \,,
\end{align}
where $\Var_\text{eq}(q_1)$ is the equilibrium variance, $\llangle q_1^4 \rrangle_\text{eq}$ is the equilibrium value of the fourth cumulant of the heat current, called kurtosis, and 
%
%%%%%%%%%%%%%%%%%%%%%%%%%%%%%%%%%%%%%%%%%%%%%%%%%%%%%%%%%%%%%%%%%%%%
\begin{figure}[t]
	\centering
	\includegraphics[width=0.9\linewidth]{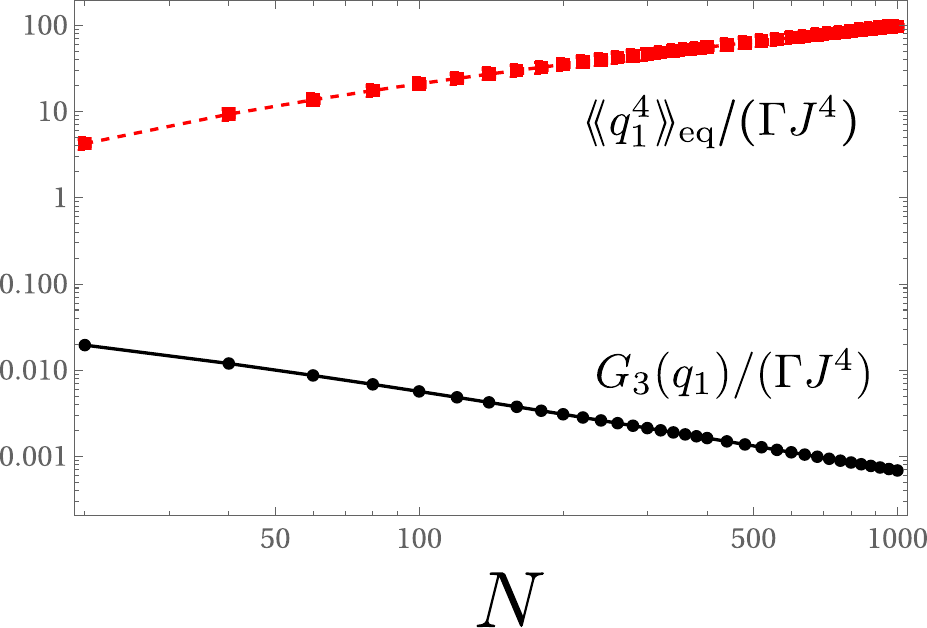}
	\caption{The finite size scaling (on a log-log scale) of the third-order response $G_3(q_1)$ and the equilibrium kurtosis $\llangle q_1^4 \rrangle_\text{eq}$ at the equilibrium phase transition point ($T_1=T_2=T_c$). Other parameters as in Fig.~\ref{fig:tempdep}. Lines added for eye guidance.}
	\label{fig:scaling-g3kurt}
\end{figure}
%%%%%%%%%%%%%%%%%%%%%%%%%%%%%%%%%%%%%%%%%%%%%%%%%%%%%%%%%%%%%%%%%%%%	
\begin{align}
G_3(q_1) = \left( \frac{\partial^3 \langle \dot{q}_1 \rangle}{\partial \Delta \beta^3} \right)_{\Delta \beta=0}
\end{align}	
is the third-order response of the heat current at equilibrium. It can be calculated using the approach presented in Sec.~\ref{subsec:curresp}. Here, due to system symmetry, the current variance depends only on even powers of $\Delta \beta$. From Fig. \ref{fig:scaling-g3kurt}, we see that $G(q_3)$ vanishes as a power law with system size, while $\llangle q_1^4 \rrangle_\text{eq}$ diverges as a power law. Thus, the heat current variance consists of two contributions: the equilibrium contribution, which vanishes as a power law, and the nonequilibrium contribution, which diverges as a power law. The latter is mainly related to divergent equilibrium kurtosis. The nonmonotonic behavior of the current variance for small temperature biases is thus a result of competition of the vanishing equilibrium contribution and the divergent nonequilibrium contribution, with the latter becoming dominant for large system sizes.

Qualitatively, we can relate the divergence of the nonequilibrium contribution to a competition of two baths, with the cold one tending to relax the system to the magnetically ordered state, and the hot bath tending to disorder the system. At the same time, we can see that a similar frustration between the ordering and disordering dynamics is already present at equilibrium and is revealed by the behavior of equilibrium kurtosis. This illustrates the usefulness of higher cumulants in providing insight into the dynamics of open systems~\cite{levitov2004counting,cuevas2003full,braggio2011superconducting,wang2007full,urban2008coulomb,ho2019counting,barato2015skewness,wampler2021skewness,ptaszynski2022bounds,gerry2023random,reulet2003environmental,gabelli2009full,pinsolle2018non,delvenne2023bounding,delvenne2024moments}.

%%%%%%%%%%%%%%%%%%%%%%%%%%%%%%%%%%%%%%%%%%%%%%%%%%%%%%%%%%%%%%%%%%%%
\begin{figure}
	\centering
	\includegraphics[width=0.9\linewidth]{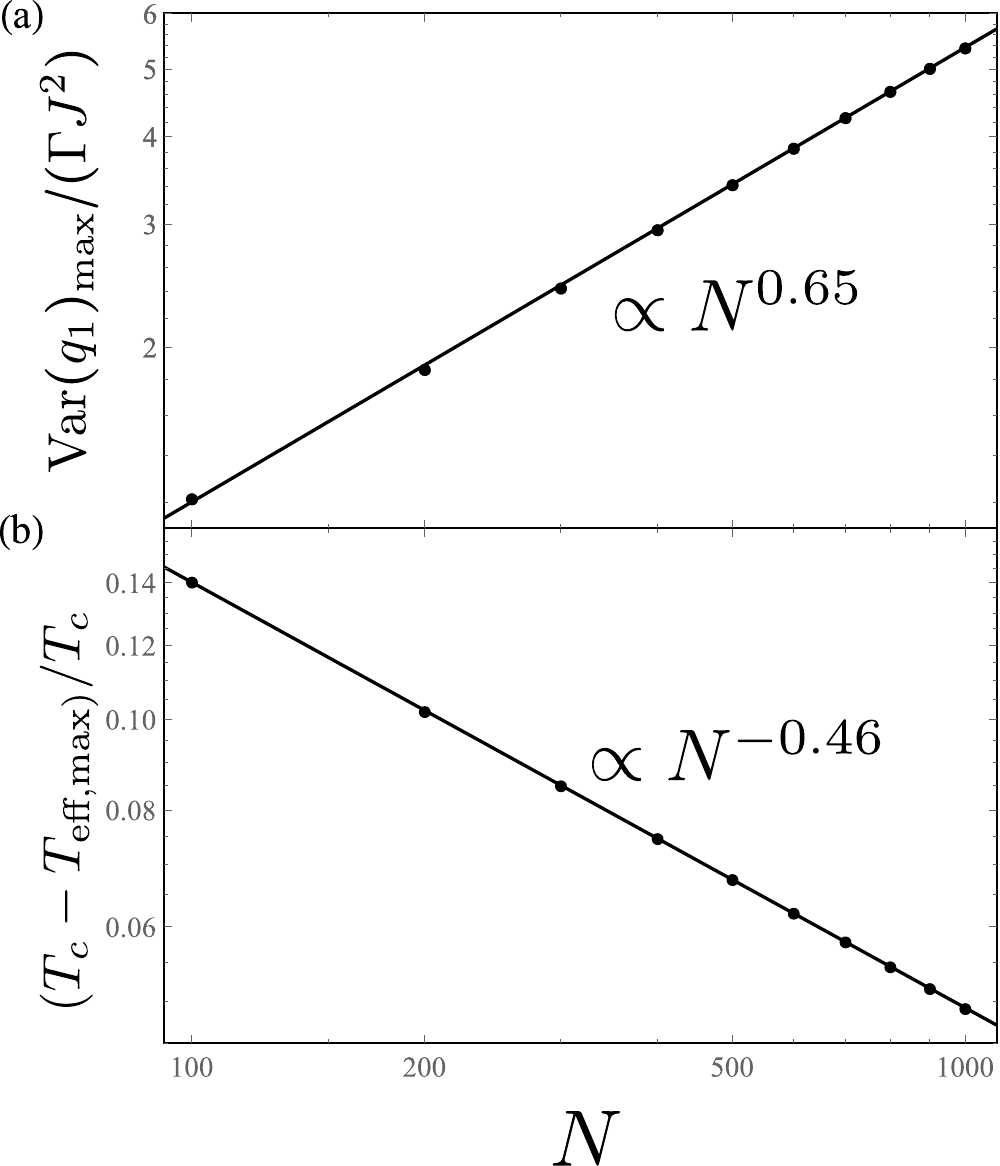}
	\caption{The finite-size scaling of the magnitude of the peak of the heat current variance (a) and the displacement of the peak position, labeled $T_\text{eff,max}$, from $T_c$ (b). Parameters as in Fig.~\ref{fig:tempdep}~(b). The figure is plotted on the log-log scale. The black dots represent the exact results, while the solid lines represent the fitted power-law behavior.}
	\label{fig:peak-tempdep}
\end{figure}
%%%%%%%%%%%%%%%%%%%%%%%%%%%%%%%%%%%%%%%%%%%%%%%%%%%%%%%%%%%%%%%%%%%%	
Finally, to complete our analysis, let us analyze the behavior of the peak of current fluctuations in their dependence on the effective temperature (see the analogous procedure in Ref.~\cite{nguyen2018phase}). As can be observed in Fig.~\ref{fig:tempdep}, the maximum of that peak occurs at effective temperatures slightly lower than $T_c$. The magnitude (height) of that peak, as well as the displacement of the maximum of that peak from $T_c$, are plotted in Fig.~\ref{fig:peak-tempdep} as a function of system size. As shown, the peak magnitude increases polynomially with system size, with the scaling exponent estimated as $0.65$. The displacement of the peak position from $T_c$ decays with system size, which can also be approximated by a power law, with the scaling exponent estimated as $-0.46$.

\subsection{Critical isotherm} \label{subsec:critical-isotherm}
%%%%%%%%%%%%%%%%%%%%%%%%%%%%%%%%%%%%%%%%%%%%%%%%%%%%%%%%%%%%%%%%%%%%
\begin{figure}
	\centering
	\includegraphics[width=0.9\linewidth]{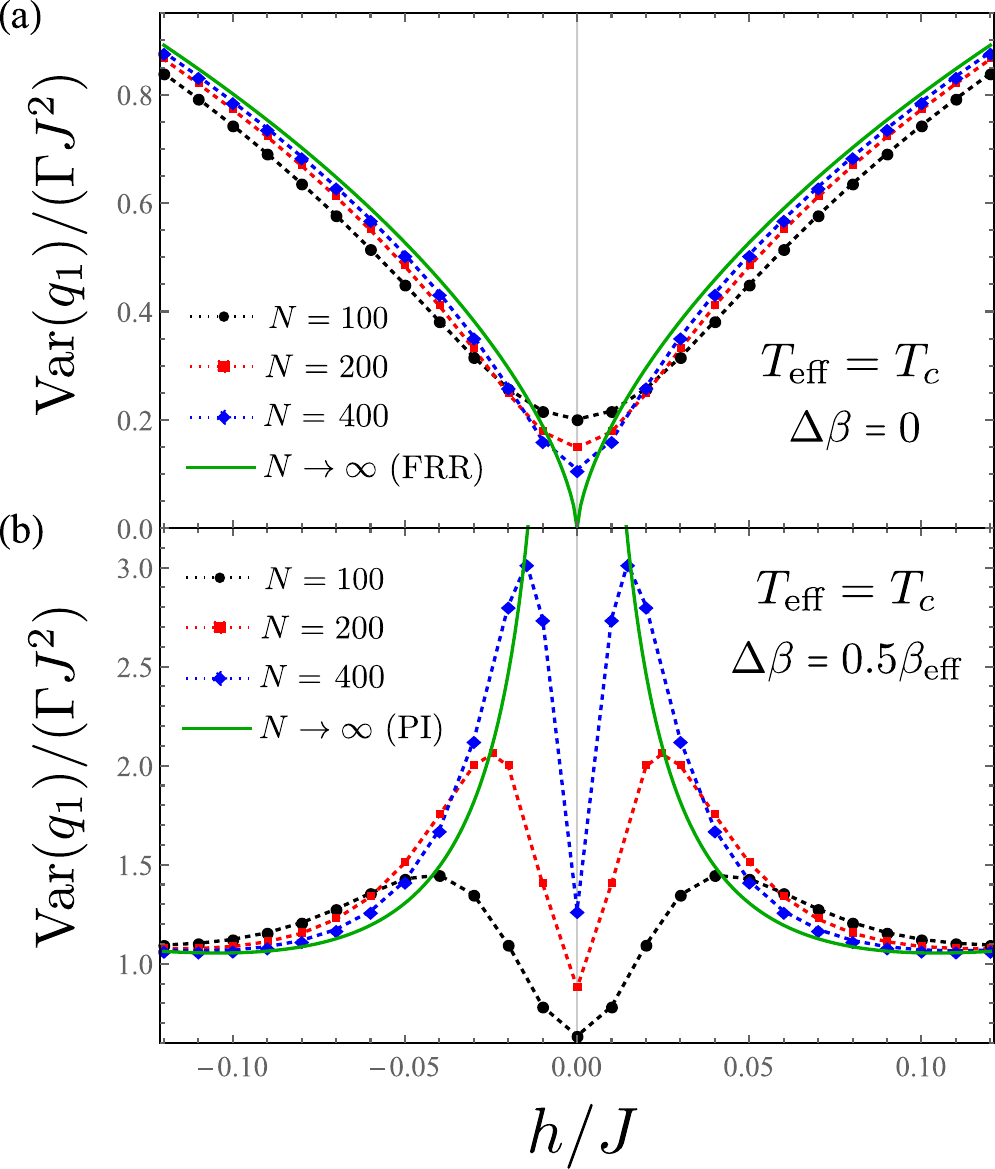}
	\caption{The magnetic field dependence of the heat current variance $\Var(q_1)$ for the effective temperature $T_\text{eff}=T_c$ and different system sizes $N$ at equilibrium (a) and for $\Delta \beta=0.5 \beta_\text{eff}$ (b). The green solid line in (a) represents the equilibrium fluctuation-response relations (FRR), Eq.~\eqref{eq:eqfluctsym}, while in (b) it represents the path integral (PI) results. The finite-size results are represented by large dots, and the dotted lines are added for eye guidance. Parameters: $\Gamma_1=\Gamma_2$, $\beta_1=\beta_\text{eff}-\Delta \beta/2$, $\beta_2=\beta_\text{eff}+\Delta \beta/2$.}
	\label{fig:hdep-isotherm}
\end{figure}
%%%%%%%%%%%%%%%%%%%%%%%%%%%%%%%%%%%%%%%%%%%%%%%%%%%%%%%%%%%%%%%%%%%%	
%%%%%%%%%%%%%%%%%%%%%%%%%%%%%%%%%%%%%%%%%%%%%%%%%%%%%%%%%%%%%%%%%%%%
%%%%%%%%%%%%%%%%%%%%%%%%%%%%%%%%%%%%%%%%%%%%%%%%%%%%%%%%%%%%%%%%%%%%
\begin{figure}
	\centering
	\includegraphics[width=0.9\linewidth]{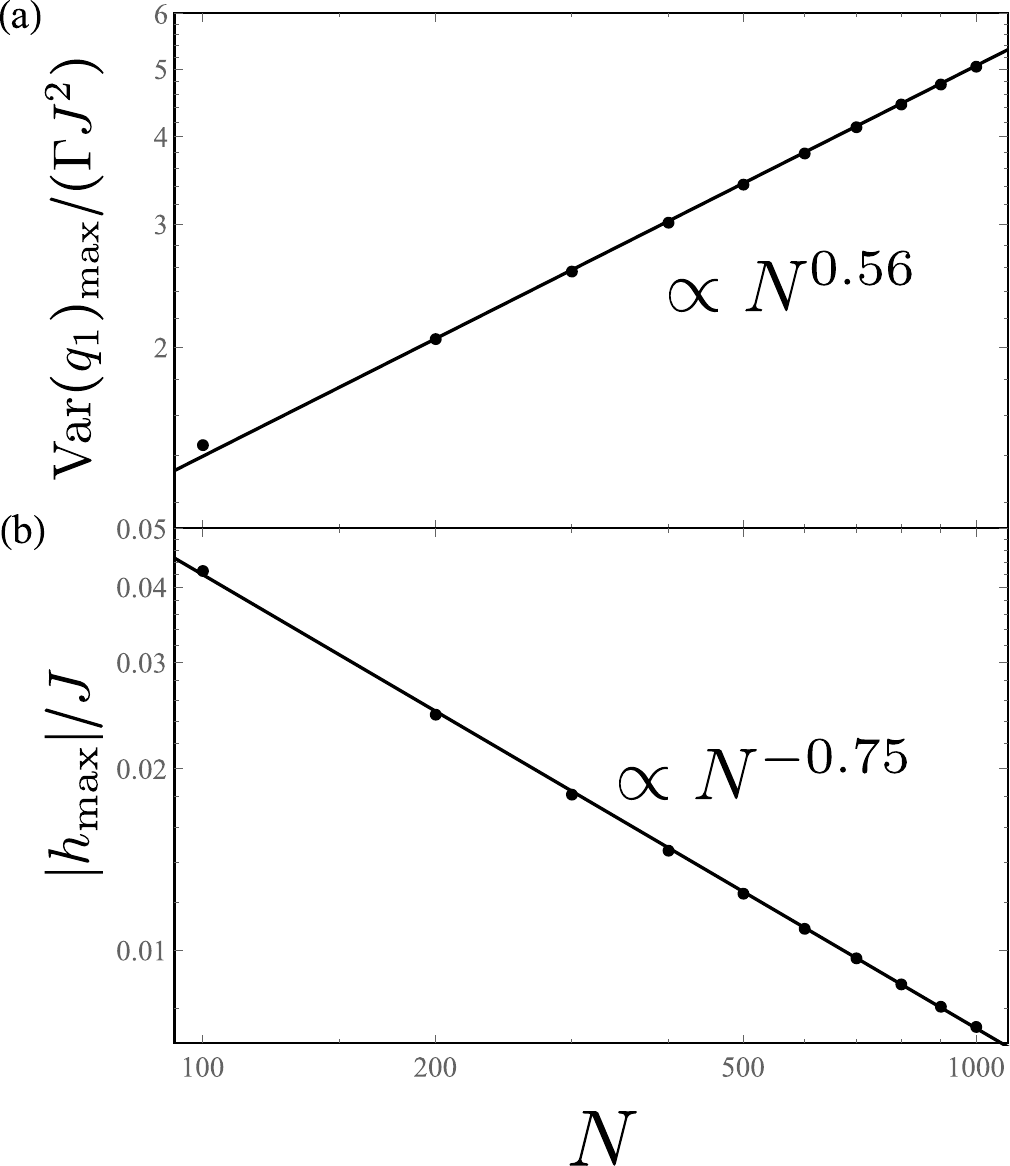}
	\caption{The finite-size scaling of the magnitude of the peak of the heat current variance (a) and the peak position (b) for parameters as in Fig.~\ref{fig:hdep-isotherm}~(b). The figures are plotted on the log-log scale. The black dots represent the exact results, while the solid lines represent the fitted power-law behavior.}
	\label{fig:peak-isotherm}
\end{figure}
%%%%%%%%%%%%%%%%%%%%%%%%%%%%%%%%%%%%%%%%%%%%%%%%%%%%%%%%%%%%%%%%%%%%	

In the previous section, we considered the situation when the critical point (i.e., the continuous phase transition point) $T_\text{eff}=T_c$, $h=0$ is crossed by sweeping the effective temperature. However, alternatively, this point can be crossed by sweeping the magnetic field along the \textit{critical isotherm}, that is, for a constant $T_\text{eff}=T_c$. The magnetic field dependence of the heat current fluctuations in such a case is presented in Fig.~\ref{fig:hdep-isotherm}. First, we notice that, both in and out of equilibrium, $\Var(q_1)$ is symmetric with respect to $h=0$. This is related to the symmetry of the model with respect to the simultaneous reversal of magnetization and magnetic field ($M \rightarrow -M$, $h \rightarrow -h$). Second, in equilibrium [Fig. \ref{fig:hdep-isotherm}~(a)], the current fluctuations exhibit a dip at $h=0$, which deepens with system size. This agrees with the predictions of the fluctuation-response relation~\eqref{eq:eqfluctsym} for $N \rightarrow \infty$, according to which the noise vanishes at $h=0$ and increases with $|h|$. Further, one can observe that for $N \rightarrow \infty$, current fluctuations exhibit a nonanalytic behavior at $h=0$ (specifically, exhibit a kink), which is analogous to the behavior at $T_\text{eff}=T_c$ in the temperature dependence. This can be explained by a nonanalytic behavior of magnetization as a function of the magnetic field at the critical point, which behaves as $m_0 \propto \text{sgn}(h)|h|^{1/3}$~\cite{kochmanski2013curie}. Inserting this into Eq.~\eqref{eq:eqfluctsym}, one finds that the heat current variance behaves close to $h=0$ as $\Var(q_\alpha) \propto |h|^{2/3}$.

In contrast, out of equilibrium [Fig. \ref{fig:hdep-isotherm}~(b)], the current fluctuations exhibit two pronounced peaks that are mirror symmetric with respect to $h=0$. As further demonstrated in Fig.~\ref{fig:peak-isotherm}, analogously to the behavior of the noise peak in the effective temperature dependence (Fig.~\ref{fig:peak-tempdep}), the peak magnitude (resp.\ peak position) increases (resp.\ decays) polynomially with system size. This agrees with the predictions of the path integral approach, according to which the current fluctuations diverge at $h=0$ in the thermodynamic limit.

We notice that the observed behavior of the heat current fluctuations out of equilibrium is slightly different from the behavior of the entropy production fluctuations along a critical line in the Schl\"{o}gl model (which can be regarded as analogous to the critical isotherm) analyzed in Ref.~\cite{remlein2024nonequilibrium}. In that case, the fluctuations exhibit only a single peak near a critical point. This may be related to the fact that the Schl\"{o}gl model does not possess any symmetry that is analogous to the symmetry of the Curie-Weiss model with respect to magnetization and field reversal.

\subsection{Field-driven transition} \label{subsec:field}
%%%%%%%%%%%%%%%%%%%%%%%%%%%%%%%%%%%%%%%%%%%%%%%%%%%%%%%%%%%%%%%%%%%%
\begin{figure}
	\centering
	\includegraphics[width=0.9\linewidth]{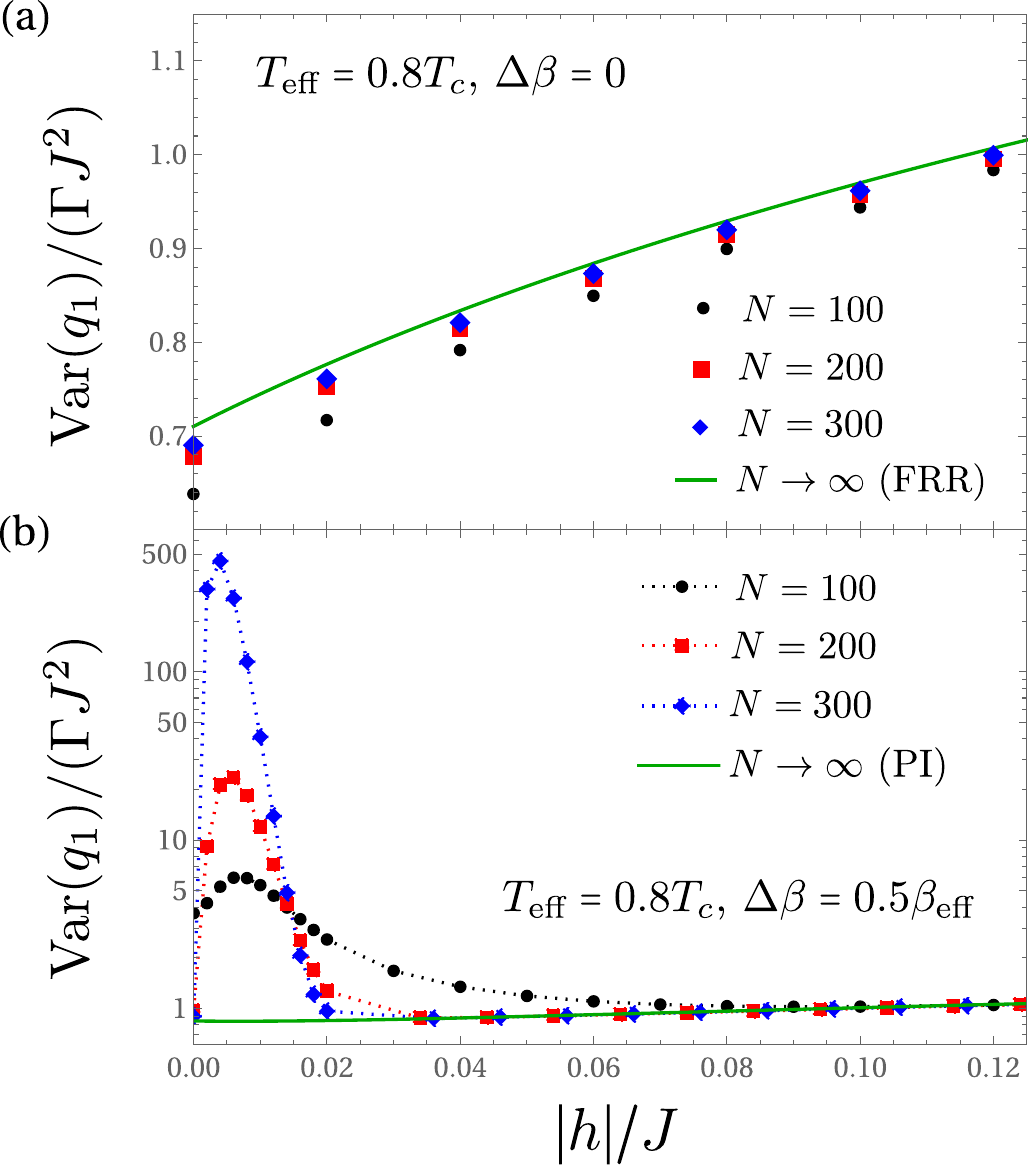}
	\caption{The magnetic field dependence of the heat current variance $\Var(q_1)$ for the effective temperature $T_\text{eff}=0.8T_c$ and different system sizes $N$ at equilibrium (a) and for $\Delta \beta=0.5 \beta_\text{eff}$ (b). The green solid line in (a) represents the equilibrium fluctuation-response relations (FRR), Eq.~\eqref{eq:eqfluctsym}, while in (b) it represents the path integral (PI) results. (b) is plotted on the logarithmic scale, and the dotted lines are added for eye guidance. Parameters: $\Gamma_1=\Gamma_2$, $\beta_1=\beta_\text{eff}-\Delta \beta/2$, $\beta_2=\beta_\text{eff}+\Delta \beta/2$.}
	\label{fig:hdep}
\end{figure}
%%%%%%%%%%%%%%%%%%%%%%%%%%%%%%%%%%%%%%%%%%%%%%%%%%%%%%%%%%%%%%%%%%%%	
%%%%%%%%%%%%%%%%%%%%%%%%%%%%%%%%%%%%%%%%%%%%%%%%%%%%%%%%%%%%%%%%%%%%
\begin{figure}
	\centering
	\includegraphics[width=0.9\linewidth]{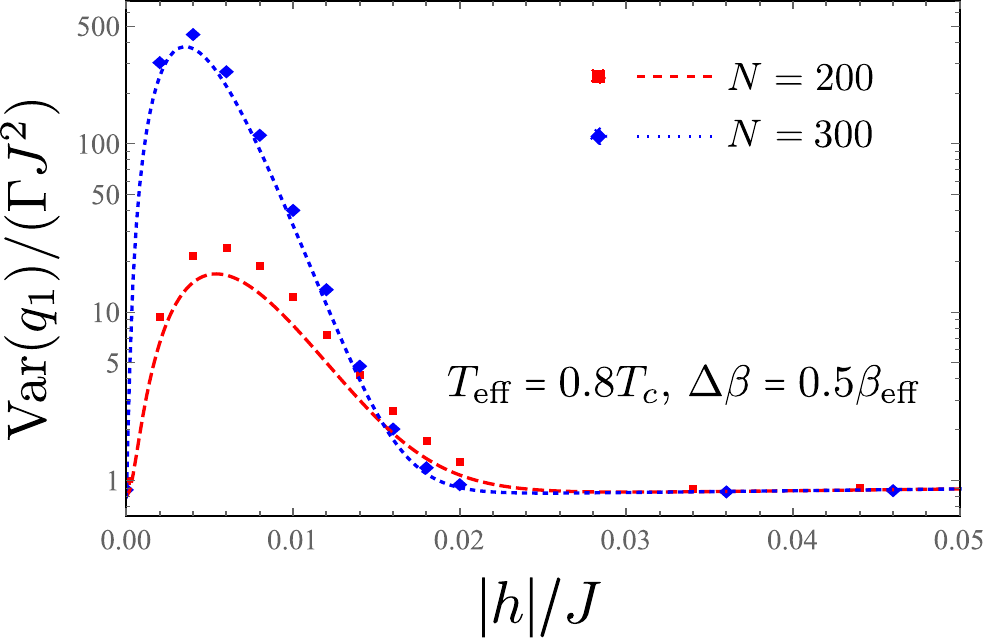}
	\caption{The magnetic field dependence of the heat current variance $\Var(q_1)$, plotted on the logarithmic scale, for two different system sizes $N$. Dots represent the exact results while lines represent the sum of current variances calculated using the path integral approach and the two-state model. Parameters as in Fig.~\ref{fig:hdep}~(b).}
	\label{fig:hdep-sum}
\end{figure}
%%%%%%%%%%%%%%%%%%%%%%%%%%%%%%%%%%%%%%%%%%%%%%%%%%%%%%%%%%%%%%%%%%%%	
Let us now turn our attention to the behavior of current fluctuations at the magnetic-field-driven transition occurring for $h=0$ and $T_\text{eff}<T_c$. In Fig.~\ref{fig:hdep} we present the magnetic field dependence of the heat current variance at equilibrium (a) and for $\Delta \beta=0.5 \beta_\text{eff}$ (b). Since the current variance is symmetric with respect to $h=0$ (see Sec.~\ref{subsec:critical-isotherm}), for better visibility, we now plot the results as a function of $|h|$. At equilibrium, the finite-size results agree well with the fluctuation-response relation. As at the critical isotherm, in the thermodynamic limit, the heat current variance behaves nonanalytically (exhibits a kink) at $h=0$, witnessing the presence of phase transition. In contrast, out of equilibrium, the finite-size results for the current variance develop a strong peak for finite values of $|h|$ near the phase transition point. The path integral approach does not capture the presence of this peak, though it well describes the heat current fluctuations for a sufficiently large magnetic field. The reason is that now this peak originates from stochastic switching between two current values associated with different fixed points of the system, namely, the absolutely stable and metastable magnetization states. This phenomenon often (though not always~\cite{ptaszynski2024finite}) occurs in systems that undergo discontinuous phase transitions~\cite{nguyen2020exponential,remlein2024nonequilibrium,fiore2021current}.
	
As discussed in Sec.~\ref{subsec:tsm}, the contribution to noise originating from stochastic switching between fixed points can be approximately described using an effective two-state model. This suggests that we can approximate noise in the whole range of $h$ by adding this contribution to path integral results. We verify this in Fig.~\ref{fig:hdep-sum}; here we present a smaller range of $h$ than in Fig.~\ref{fig:hdep} to better visualize the noise peak. As shown, the applied approximation reproduces well the noise behavior, with the quantitative agreement improving for large $N$.

%%%%%%%%%%%%%%%%%%%%%%%%%%%%%%%%%%%%%%%%%%%%%%%%%%%%%%%%%%%%%%%%%%%%
\begin{figure}[b]
	\centering
	\includegraphics[width=0.9\linewidth]{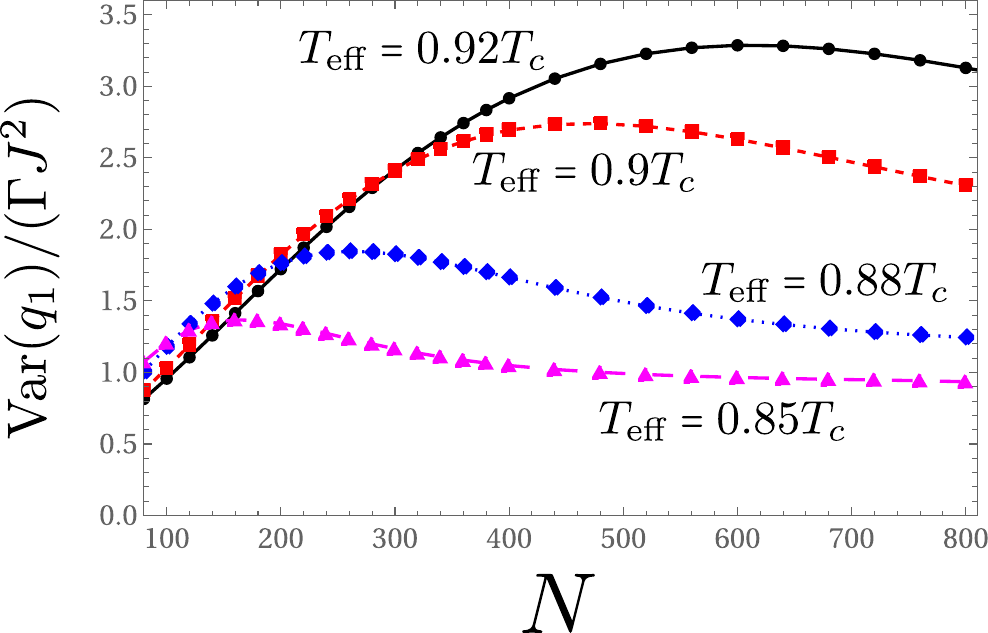}
	\caption{The finite-size scaling of the heat current variance $\Var(q_1)$ at the phase transition point ($h=0$) for different effective temperatures $T_\text{eff}$ and $\Delta \beta=0.4 \beta_\text{eff}$. Other parameters as in Fig.~\ref{fig:hdep}. Lines added for eye guidance.}
	\label{fig:scaling-var-field}
\end{figure}
%%%%%%%%%%%%%%%%%%%%%%%%%%%%%%%%%%%%%%%%%%%%%%%%%%%%%%%%%%%%%%%%%%%%		

At first glance, the behavior of the heat current fluctuations out of equilibrium (the presence of the noise peaks) may appear to be similar to the one observed at the critical isotherm (Sec.~\ref{subsec:critical-isotherm}). However, there are crucial differences. First, as shown in Fig.~\ref{fig:scaling-var-field}, the current variance does not diverge with system size when evaluated exactly at the phase transition point ($h=0$). This differs from the behavior at the critical point (continuous phase transition) where the fluctuations diverge polynomially (Fig.~\ref{fig:scaling-var-temp}). Instead, fluctuations exhibit nonmonotonic behavior: they first increase but later decrease with $N$, tending to saturate at some finite value. This saturation value is higher for effective temperatures closer to $T_c$, where fluctuations diverge. We notice that this behavior of fluctuations differs significantly from previously considered models of phase transitions, where current fluctuations diverged either polynomially~\cite{nguyen2018phase,oberreiter2021stochastic,remlein2024nonequilibrium,kewming2022diverging} or exponentially~\cite{nguyen2020exponential,remlein2024nonequilibrium,kewming2022diverging,fiore2021current} at the phase transition point. In particular, even though the system exhibits a stochastic switching between fixed points, this does not lead to noise enhancement. This can be explained using the two-state model: for $h=0$, both fixed points are associated with the same heat current value, and thus the prefactor $\langle \dot{q}_1 \rangle_+-\langle \dot{q}_1 \rangle_-$, appearing in formula~\eqref{eq:vartsm} for the current variance, disappears.

%%%%%%%%%%%%%%%%%%%%%%%%%%%%%%%%%%%%%%%%%%%%%%%%%%%%%%%%%%%%%%%%%%%%
\begin{figure}
	\centering
	\includegraphics[width=0.9\linewidth]{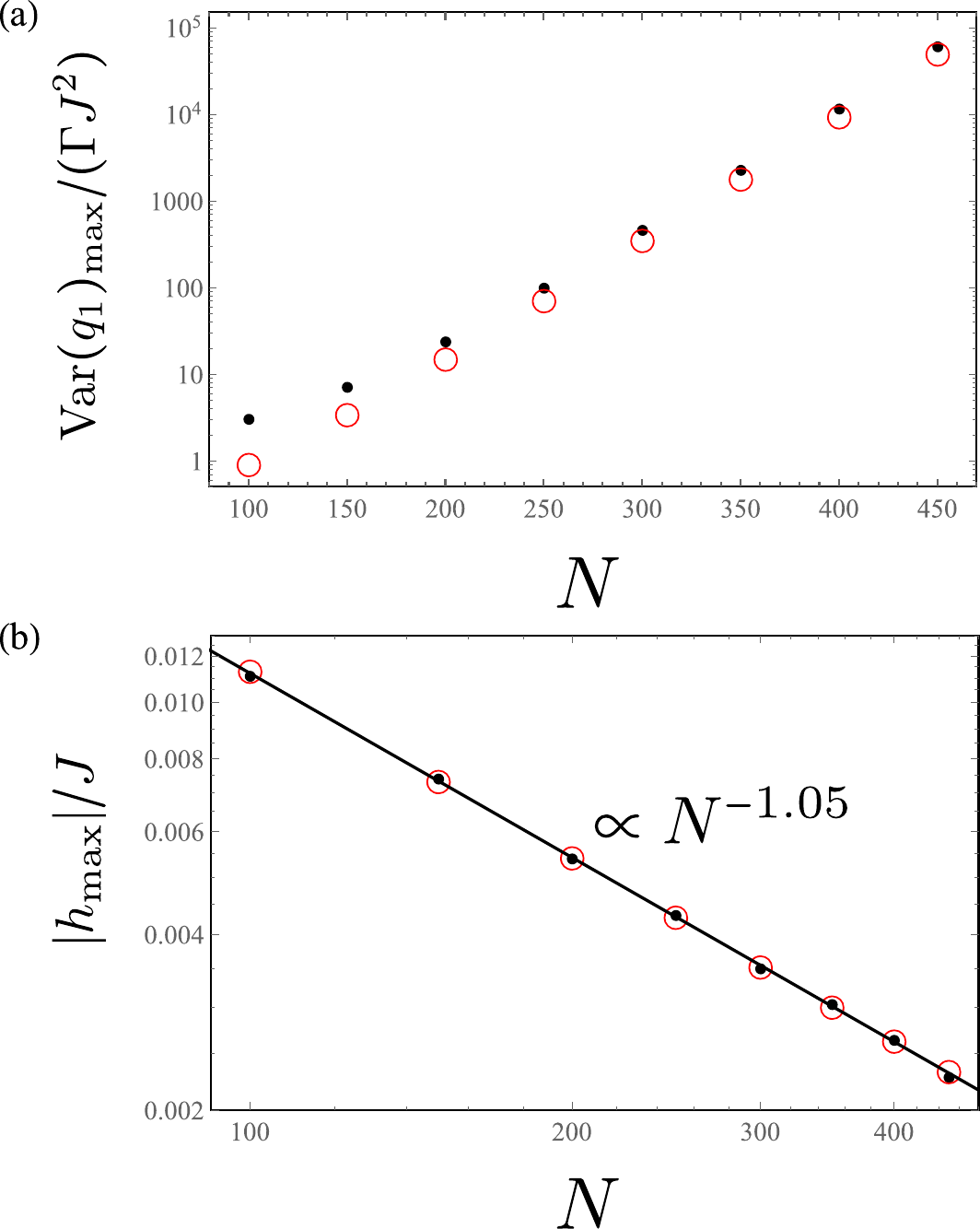}
	\caption{The finite-size scaling of the magnitude of the peak of the heat current variance (a) and the peak position (b) for parameters as in Fig.~\ref{fig:hdep}~(b). The black dots represent the exact results, while the red circles represent predictions of the two-state model. (a) is plotted on the logarithmic scale and (b) is plotted on the log-log scale. The solid line in (b) represents the fitted power-law behavior.}
	\label{fig:peak}
\end{figure}
%%%%%%%%%%%%%%%%%%%%%%%%%%%%%%%%%%%%%%%%%%%%%%%%%%%%%%%%%%%%%%%%%%%%	
Second, as shown in Fig.~\ref{fig:peak}~(a)\footnote{Here we focus on $N \leq 450$, since for larger systems the numerical calculations in the bistable regime become unreliable due to the small value of the spectral gap.}, the magnitude of the peak of current fluctuations diverges exponentially with system size (rather than polynomially, as is in the case of peak at the critical isotherm). At the same time, the position of this peak [Fig.~\ref{fig:peak}~(b)] shifts toward the phase transition point ($h=0$), following a power-law behavior that is close to a hyperbolic one. This behavior can be explained using the two-state model (see red circles in Fig.~\ref{fig:peak}), which reproduces well the exact results. Within this model, exponential scaling of the peak magnitude is the result of the exponential suppression of transition rates $r_\pm$, see Eq.~\eqref{eq:ratestsm}. At the same time, the shift of the peak position is the result of the competition of two factors: On the one hand, by increasing $|h|$, one magnifies the difference between the heat current values for the stationary and metastable states, $\langle \dot{q}_1 \rangle_+-\langle \dot{q}_1 \rangle_-$, which enhances the fluctuations. On the other hand,  this also magnifies the difference between quasipotential values of magnetization states $|V(m_+)-V(m_-)|$. By virtue of the large deviation principle~\eqref{eq:largedev}, this speeds up the exponential decay of the probability of the metastable state, $\min \{p_+,p_- \}$.  This reduces noise, which is proportional to the product $p_+ p_-$ [Eq.~\eqref{eq:vartsm}]. The latter effect becomes dominant for large $N$, leading to the shift of the peak position towards $h=0$. 

This leads to a somewhat paradoxical conclusion: while fluctuations do not diverge with system size when evaluated exactly at the phase transition point (Fig.~\ref{fig:scaling-var-field}), the observed peak behavior suggests that in the thermodynamic limit $N \rightarrow \infty$ they diverge ``infinitely closely'' to the phase transition point. Thus, there is certain subtlety in applying the thermodynamic limit when considering current fluctuations.\footnote{We note that a similar subtlety arises already when considering the equilibrium magnetization for the Curie-Weiss model: it is equal to 0 when the limit $h \rightarrow 0^+$ is applied before the thermodynamic limit $N \rightarrow \infty$, while it is finite when the limits are applied in the opposite order.} This also suggests that to fully account for the effect of phase transitions on current fluctuations, one needs to analyze their behavior not only at the transition point itself but also in a finite region around the phase transition point.

\subsection{Magnetic field dependence of fluctuations above the critical temperature}
%%%%%%%%%%%%%%%%%%%%%%%%%%%%%%%%%%%%%%%%%%%%%%%%%%%%%%%%%%%%%%%%%%%%
\begin{figure}[t]
	\centering
	\includegraphics[width=0.9\linewidth]{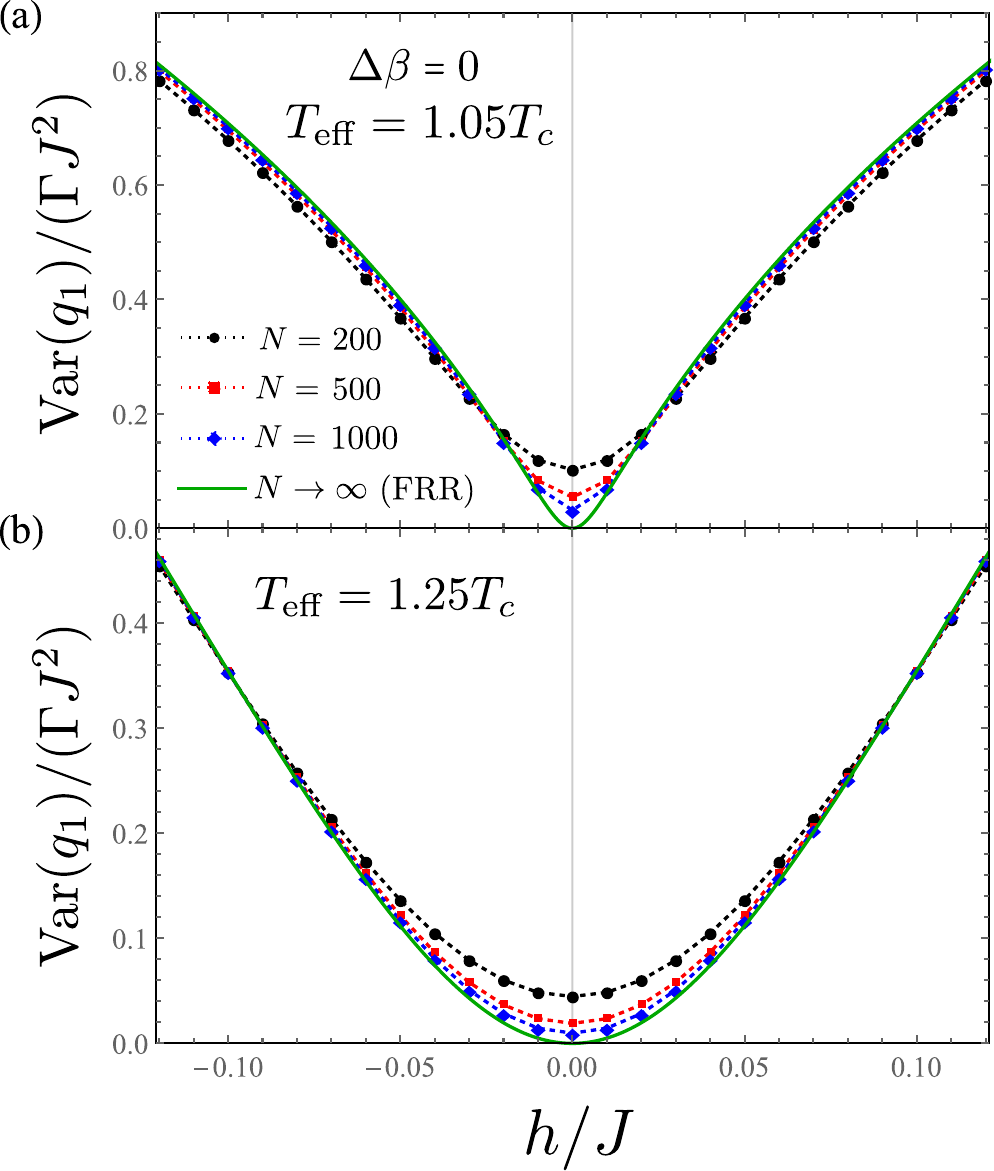}
	\caption{The magnetic field dependence of the heat current variance $\Var(q_1)$ at equilibrium ($\Delta \beta=0$) for $T_\text{eff}=1.05 T_c$ (a) and $T_\text{eff}=1.25 T_c$ (b). The green solid line in represents the equilibrium fluctuation-response relations (FRR), Eq.~\eqref{eq:eqfluctsym}. The finite-size results are represented by large dots, and the dotted lines are added for eye guidance. Parameters: $\Gamma_1=\Gamma_2$, $\beta_1=\beta_\text{eff}-\Delta \beta/2$, $\beta_2=\beta_\text{eff}+\Delta \beta/2$.}
	\label{fig:hdep-abovetc-equilibrium}
\end{figure}
%%%%%%%%%%%%%%%%%%%%%%%%%%%%%%%%%%%%%%%%%%%%%%%%%%%%%%%%%%%%%%%%%%%%
\begin{figure}
	\centering
	\includegraphics[width=0.9\linewidth]{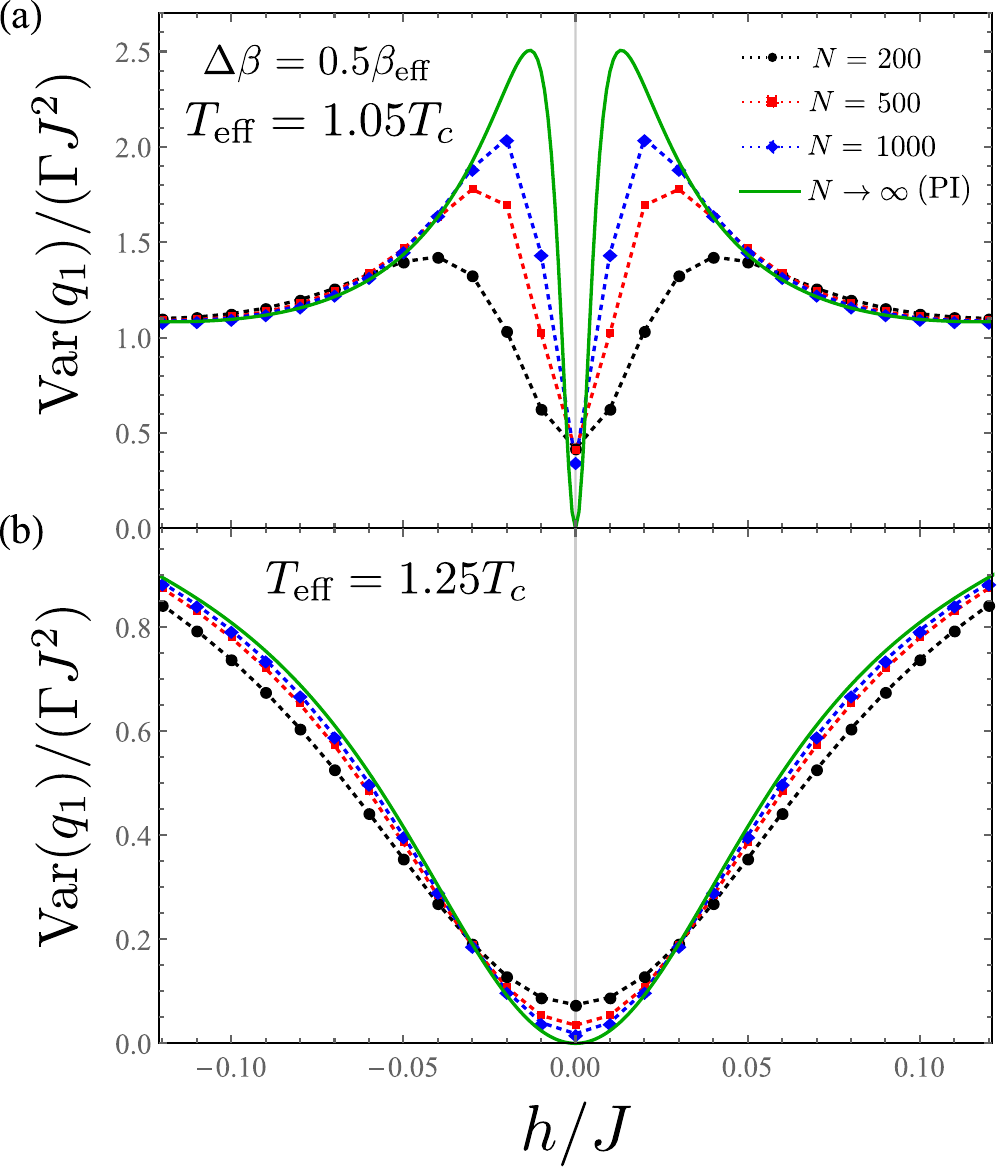}
	\caption{The same plot as in Fig.~\ref{fig:hdep-abovetc-equilibrium} but for the nonequilibrium case ($\Delta \beta=0.5 \beta_\text{eff}$). The green solid line represents the path integral (PI) results.}
	\label{fig:hdep-abovetc}
\end{figure}
%%%%%%%%%%%%%%%%%%%%%%%%%%%%%%%%%%%%%%%%%%%%%%%%%%%%%%%%%%%%%%%%%%%%	
Finally, for the sake of completeness, let us analyze the magnetic field dependence of the heat current fluctuations for $T_\text{eff}>T_c$, where no phase transition occurs. In Figs.~\ref{fig:hdep-abovetc-equilibrium} and~\ref{fig:hdep-abovetc}, we plot the results for the equilibrium ($\Delta \beta=0$) and the nonequilibrium ($\Delta \beta=0.5 \beta_\text{eff}$) cases, respectively. In both cases, we consider two effective temperatures that are close to ($T_\text{eff}=1.05 T_c$) or away from ($T_\text{eff}=1.25 T_c$) the critical temperature. As shown, both in and out of equilibrium, the heat current variance exhibits a dip at $h=0$ and is an analytic function of $h$ in the thermodynamic limit $N \rightarrow \infty$. However, when the system is out of equilibrium and close to critical temperature ($T_\text{eff}=1.05 T_c$), the system exhibits two noise peaks at the magnetic field close to $h=0$ [Fig.~\ref{fig:hdep-abovetc}~(a)], which are not observed otherwise. As predicted by the path integral approach, in contrast to the behavior for $T_\text{eff} \leq T_c$, these peaks have a finite magnitude in the thermodynamic limit. Thus, their presence might be regarded as a precursor of the divergent behavior of fluctuations observed for $T_\text{eff} \leq T_c$.

\section{Conclusions} \label{sec:concl}
Our objective was to verify whether current fluctuations always exhibit divergent behavior at the phase transition point. To this end, we studied the heat current fluctuations in the Curie--Weiss model attached to two thermal baths with different temperatures.  First, we considered the temperature-driven phase transition, which is continuous for both the magnetization and the heat current. We found that in this case the fluctuations consist of two components: the equilibrium one, which vanishes in the thermodynamic limit, and the nonequilibrium one, which exhibits a power-law divergence with system size. For a small temperature bias, this leads to a nonmonotonic scaling of fluctuations: they first decrease due to vanishing of the equilibrium contribution and then increase due to the divergence of the nonequilibrium noise component.  Qualitatively, we can relate the different behavior of equilibrium and nonequilibrium fluctuations to the competition between the ordering and disordering dynamics induced by the cold and the hot bath, respectively. At the same time, we note that a similar frustration mechanism is already present at equilibrium and manifests itself in the divergent behavior of the heat current kurtosis, that is, the fourth current cumulant. This illustrates the value of higher-order cumulants in providing insight into the dynamical behavior of open systems~\cite{levitov2004counting,cuevas2003full,braggio2011superconducting,wang2007full,urban2008coulomb,ho2019counting,barato2015skewness,wampler2021skewness,ptaszynski2022bounds,gerry2023random,reulet2003environmental,gabelli2009full,pinsolle2018non,delvenne2023bounding,delvenne2024moments}. It may be further worth investigating whether the divergent (resp.\ nondivergent) behavior of heat current fluctuations at nonequilibrium (resp.\ equilibrium) phase transition is related to the similar behavior of nonequilibrium (resp.\ equilibrium) heat capacity, reported in Ref.~\cite{beyen2024phase}.

In the next step, we analyzed the magnetic-field-driven transition that occurs below the critical temperature. At this transition, magnetization jumps discontinuously, while the heat current is continuous but exhibits a kink. In this case, the current fluctuation behavior is somewhat more subtle than in the previously investigated models~\cite{remlein2024nonequilibrium,kewming2022diverging,nguyen2020exponential,nguyen2018phase,oberreiter2021stochastic,fiore2021current}. When fluctuations are evaluated exactly at the phase transition point, they do not diverge with system size, but rather saturate at some finite value. However, one can observe the emergence of maxima in the magnetic field dependence of noise, that occur for small but finite field values. The magnitude of those maxima diverges exponentially with system size, while their positions shift towards the phase transition point. This effect is the result of stochastic switching between metastable current values, a phenomenon often~\cite{nguyen2020exponential, fiore2021current, kewming2022diverging} (though not always~\cite{ptaszynski2024finite}) occurring for discontinuous nonequilibrium phase transitions. This means that to fully account for the effect of phase transition on current fluctuations, one needs to examine their behavior not only at the phase transition point itself but also in a finite region around the phase transition point. Furthermore, we emphasize that the noise maxima grow exponentially despite the fact that the heat current is continuous at the phase transition. This makes the previously suggested picture, in which continuous (resp.\ discontinuous) phase transitions are associated with power-law (resp.\ exponential) divergence of fluctuations~\cite{nguyen2018phase,nguyen2020exponential}, more nuanced.

On the methodological side, the paper demonstrates that current fluctuations in large systems can be effectively characterized by combining two complementary approaches: The recently developed path integral approach~\cite{lazarescu2019large,herpich2020njp,gopal2022large,vroylandt2020efficiency}, which characterizes conditional current fluctuations at a specific fixed point, and the two-state model~\cite{nguyen2020exponential}, which accounts for stochastic switching between fixed points. We also note that conditional current fluctuations, considered on their own, have recently attracted some attention in the literature~\cite{fiore2021current,vroylandt2020efficiency}.

\acknowledgments

The authors thank Ashwin Gopal for useful comments on the path integral approach to current fluctuations. K.P.\ acknowledges the financial support of the National Science Centre, Poland, under the project No.\ 2023/51/D/ST3/01203.

\section*{Data availability}
Wolfram Mathematica notebooks used to obtain the numerical results and the data used in the figures are available at the following DOI: 10.5281/zenodo.14213010.

\bibliography{bibliography}	
	
\end{document}